\documentclass[11pt,preprint,flushrt]{aastex}

\usepackage{natbib,graphicx,amsmath,subfigure,color}
\def\eqq#1{Equation~(\ref{#1})}

\newcommand\eg{{\it e.g.\/}}
\newcommand\ie{{\it i.e.\/}}

\newcommand{\vecD}{\mbox{\boldmath $D$}}
\newcommand{\vecF}{\mbox{\boldmath $F$}}
\newcommand{\vecX}{\mbox{\boldmath $X$}}
\newcommand{\vecg}{\mbox{\boldmath $g$}}

\newcommand{\vecM}{\mbox{\boldmath $M$}}
\newcommand{\vecN}{\mbox{\boldmath $N$}}
\newcommand{\vecQ}{\mbox{\boldmath $Q$}}
\newcommand{\vecZ}{\mbox{\boldmath $Z$}}
\newcommand{\vect}{\mbox{\boldmath $t$}}
\newcommand{\vecx}{\mbox{\boldmath $x$}}
\newcommand{\veck}{\mbox{\boldmath $k$}}
\newcommand{\vecv}{\mbox{\boldmath $v$}}
\newcommand{\vecu}{\mbox{\boldmath $u$}}

\newcommand{\matR}{\mbox{$\bf R$}}
\newcommand{\matC}{\mbox{$\bf C$}}
\newcommand{\matB}{\mbox{$\bf B$}}
\newcommand{\matA}{\mbox{$\bf A$}}
\newcommand{\matI}{\mbox{$\bf I$}}
\newcommand{\bnab}{\boldsymbol{\nabla}}
\newcommand{\bnabg}{\boldsymbol{\nabla_g}}

\newcommand{\likeli}{\mbox{$\mathcal{L}$}}
\newcommand{\lensfit}{{\sc lensfit}}
\newcommand{\galsim}{{\sc GalSim}}
\newcommand{\great}{{\sc great3}}

\newcommand{\photoz}{photo-$z$}
\newcommand\edit[1]{#1}
\begin{document}

\slugcomment{V2.3 27 Apr 2016 Accepted by MNRAS}

\title{An accurate and practical method for inference of weak
  gravitational lensing from galaxy images}

\author{Gary M. Bernstein}
\email{garyb@physics.upenn.edu}
\author{Robert Armstrong\altaffilmark{1}}
\email{rearmstr@gmail.com}
\author{Christina Krawiec}
\email{ckrawiec@sas.upenn.edu}
\and 
\author{Marisa C. March}
\email{mamarch@sas.upenn.edu}
\affil{Department of Physics \& Astronomy, University of Pennsylvania, 
209 S.\ 33rd St., Philadelphia, PA 19104}
\altaffiltext{1}{Department of Astrophysical Sciences, Princeton
  University, Princeton, NJ 08544}
\begin{abstract}
We demonstrate highly accurate recovery of weak gravitational lensing
shear using an implementation of the Bayesian
Fourier Domain (BFD) method proposed by
Bernstein \& Armstrong (2014, BA14), extended to correct for selection
biases.  The BFD formalism is rigorously
correct for  Nyquist-sampled, background-limited,
uncrowded image of background galaxies.
BFD does not assign shapes to galaxies, instead compressing the pixel
data \vecD\ into a vector of moments \vecM, such that we have an
analytic expression for the probability $P(\vecM|\vecg)$ of obtaining
the observations with gravitational lensing distortion \vecg\ along
the line of sight.
We implement an algorithm for conducting BFD's integrations over
the population of unlensed source galaxies which 
measures $\approx10$ galaxies/second/core with good scaling properties.
Initial tests of this code on
\edit{$\approx 10^9$ simulated lensed galaxy images recover the simulated shear
to a fractional accuracy of $m=(2.1\pm0.4)\times10^{-3},$ substantially more
accurate than has been demonstrated previously for any generally applicable
method.} Deep sky exposures generate a
sufficiently accurate approximation to the noiseless, unlensed galaxy
population distribution assumed as input to BFD.
Potential extensions of the method include simultaneous measurement of
magnification and shear; multiple-exposure, multi-band observations; and joint
inference of photometric redshifts and lensing tomography.
\end{abstract}

\section{Introduction}
Weak gravitational lensing (WL) provides an unambiguous measurement of the
second (and potentially higher) derivatives of the scalar gravitational potential
along the line of sight.  This has made WL a critical observational
window into the behavior and history of the components of
the Universe that source the gravitational potential but do not absorb or emit photons.
WL can in addition test the laws of gravitation relating the potential
to the matter.  Several visible/near-IR surveys of thousands of square
degrees of sky are now underway, with measurement of WL signals from
images of $O(10^8)$ galaxy images as a
primary goal: the \emph{Dark Energy Survey (DES)} \citep{jarvissv},
the \emph{Kilo-Degree Survey (KiDS)} \citep{kids}, and the
\emph{Hypersuprime-cam Survey (HSC)} \citep{hsc}.  Even
more ambitious 
visible/near-IR WL surveys are planned to measure $>10^9$ galaxy
images in the 2020's: \emph{the Large
  Synoptic Survey Telescope (LSST)},\footnote{{\tt
    http://www.lsst.org}} the \emph{Euclid} spacecraft \citep{Euclidredbook},\footnote{{\tt
    http://www.euclid-ec.org}}  and
the \emph{Wide-Field Infrared Survey Telescope
  (WFIRST).}\footnote{{\tt http://wfirst.gsfc.nasa.gov}}
WL distortions have also been detected in radio images of galaxies
\citep{chang,DB15} and of the cosmic microwave background radiation
\citep{spt,act,planck,descmb}.

WL signals are known to be difficult to extract from sky images.
Along nearly all lines of sight, the dominant manifestations of
lensing are a magnification $\mu$ and a shear $(g_1,g_2)$ of the
image that define a local rotation-free linear transformation of the
sky image.  In this paper we will concentrate on estimating these
parameters from real-space images of galaxies (with some mention of
interferometric imaging in Section \ref{interferometry}).  The
difficulties include: the signal is weak, with
magnification and shear having RMS amplitudes of $\approx0.02$ on
cosmological lines of sight; the source galaxies are typically of
comparable intrinsic size to the point spread function (PSF) of the
imaging, and the PSF has asymmetries and variation larger than the
lensing signal; the WL information is primarily in galaxies with
modest signal-to-noise ratios ($S/N\lesssim25$); and the light distribution from
galaxies, unlike the CMB, is not described by any known
statistical process.

A series of community-wide blind ``challenges''  in extracting shear
distortion applied to simulated sky images provides a good summary of
progress in overcome these obstacles, most recently the \great\
challenge \citep{great3}.  A common quantification of the
accuracy of inference is to compare a measured shear component $g_{\rm
  meas}$ to the value $g_{\rm true}$ inserted into the simulation by
\begin{equation}
g_{\rm meas} - g_{\rm true} = m g_{\rm true} + c,
\end{equation}
where $m$ is the multiplicative or calibration error on the shear, and
$c$ is spurious signal uncorrelated with the input shear, \eg\ due to
leakage of PSF asymmetries into $g_{\rm meas}.$  The ambitious
next-generation surveys require $|m| \lesssim 10^{-3}$ and $c_{\rm
  RMS} \lesssim 10^{-3.5}$ in order to keep shear-measurement errors
from degrading the accuracy of cosmological inferences
\citep{htbj,amararefregier}. The \great\ simulated galaxy samples are large
enough to measure $m$ to $\pm0.003$ and $c$ to $\pm10^{-4}$ at 68\%
confidence. Over 1000 measurements were submitted to \great\ using
$\approx20$ distinct methodologies.  None were consistently able to
achieve $m=c=0$ at this level of accuracy, though the {\sc
  sFIT} \edit{algorithm \citep[see][]{great3}} did so in more than one branch of the
challenge.  There has not to date been any demonstration of a
practical shear inference method that is accurate at the
part-per-thousand level we will soon want.  In this paper we show that the
Bayesian Fourier Domain (BFD) method proposed by \citet[BA14]{BA14}
is in this regime, validating the method on simulations similar to some
branches of \great.  Our validation tests are more demanding than
\great\ in the
sense that we include lower-$S/N$
source images and hence require some correction for selection
biases.\footnote{\edit{We have not applied BFD to the public \great\
  challenge data.  The underlying galaxy population for the \great\
  images was chosen to depend on the observing conditions, hence
  violating BFD's basic assumption that deep images are sampling the
  same population as the target images.}}

Most of the current effort on shear measurement is directed towards
model-fitting methods, whereby a parametric model of each galaxy's
appearance is convolved with the PSF and compared to the pixel data.
The models are usually concentric combinations of exponential and
deVaucouleurs ellipsoids (or other S\'{e}rsic profiles).
Galaxies are assigned the ellipticity of the model which has the
maximum likelihood of reproducing the pixel data, or an ellipticity is
assigned from some weighting of the likelihood surface.  Model-fitting
methods can of course accrue biases because real galaxies are not
fully described by the model \citep{vb,fdnt}.
Other methods 
\citep[\eg][]{ksb,bj02,fdnt} assign an ellipticity via
model-independent means, typically as some combination of weighted central
moments. In both approaches, a shear estimator $g_{\rm meas}$ is produced from
some weighted sum of the galaxy ellipticities.  While these methods
overcome many of the WL inference difficulties noted above, none
includes a rigorous treatment of the propagation of image noise
through the measurement, and hence incur ``noise biases,'' \eg\
because the maximum-likelihood parameters \edit{respond asymmetrically to
noise,} hence can accrue bias in the presence of noise.
Furthermore all these methods are subject to selection biases, as the
criteria for inclusion and/or weighting of galaxies'
ellipticities implicitly favor certain pre-lensing orientations of the
sources.  Recovery of unbiased WL estimators then relies upon applying
corrections derived empirically from application of the method to
simulated images with known lensing distortion
\citep[\eg][]{gruen,tomek}.  The accuracy of such 
corrections thus depends upon the simulated images capturing all
salient characteristics of the real sky \edit{and the observing process.}  

The largest recent WL cosmology surveys have adopted variants of
model-fitting.  The {\it CFHTLS} and {\it KiDS} surveys use the
\lensfit\ code, with empirical corrections averaging $m\approx0.05$
but as large as $m\approx0.20$ applied as described in
\citet{lensfitcfh}.  
The \emph{DES} Science Verification WL results \citep{jarvissv} use two
parallel codes: {\sc ngmix} \citep{ngmix} and {\sc im3shape} \citep{im3shape}, the
latter with empirical corrections for noise and selection bias which
again range up to $\approx0.2$.  \citet{jarvissv} conclude from
a battery of internal tests and inter-comparisons of the methods that an
uncertainty of $\pm0.05$ should be assigned to $m$ in the final shear
catalog.  This accuracy is shown to be sufficient for the
preliminary results from \emph{DES},  but this and other surveys in progress
will soon require WL inferences with $\approx10\times$ better accuracy.  Further,
it is disconcerting that the simulation-derived corrections are
$\approx50\times$ larger than the accuracy needed for next-generation
WL projects.

BA14 propose a different approach to shear inference.  They suggest
skipping the estimation of galaxy shape properties, instead
evaluating from the outset the probability $P(\vecD_i |
\vecg)$ that the pixel data $\vecD_i$ from galaxy $i$ would be
produced for lensing \vecg\ on its line of sight.  While a single
galaxy provides only weak discrimination on the shear in that the
derivative $\bnabg P$ is small, a large population of sources can
tightly constrain the mean \vecg, or any model for spatially varying WL.
The BA14 method relies upon having high-$S/N$ observations
of a representative sample of source galaxies, which can be obtained from
observations of a subset of the survey region with longer integration times.
\edit{In other words the model for the unlensed population of target
  galaxies is that it is drawn from the finite number of galaxies
  observed in these deep-integration fields---or more specifically,
  that the targets' \emph{moments} are drawn from those of the
  deep-field galaxies, up to rotation and parity transformations (as
  explained below).}

In Section~\ref{derivations} we re-derive the
BA14 BFD method, extending the method to include a prescription for
detection and selection of sources for which selection biases can be
calculated to high accuracy.  Section~\ref{algorithm} describes our
computational implementation of the BFD method, and in Section~\ref{validation}
we demonstrate using simulations similar to \great\ that our
implementation does indeed recover input shear near part-per-thousand
accuracy, potentially unlocking the full power of current and future
lensing surveys.

In Section~\ref{approx} we take inventory of the assumptions and
approximations made in deriving the BFD estimators, and assess the
extent to which violations of these in real data might compromise the
measurement accuracy.  We find no show-stoppers yet.  In
Section~\ref{extensions} we sketch a number of straightforward
extensions of the current BFD implementation that will extend its
science reach, \eg\ to measurements of magnification as well as shear,
to multi-band or interferometric measurements, and to integrating
photometric redshift and tomographic WL inference into a single
measurement process.

\section{Formalism}
\label{derivations}
Our goal is to infer the lensing distortion \vecg\ from the
observational data vector \vecD.  Our current implementation assumes
pure-shear distortions, so $\vecg=(g_1,g_2)$, but the formalism is
unchanged if we include magnification $\mu$ in \vecg\ as well.  By
Bayes' theorem
\begin{equation}
P(\vecg | \vecD) = \frac{P(\vecD | \vecg) P(\vecg)}{P(\vecD)}.
\end{equation}
We will not be concerned with the normalization by the
evidence $P(\vecD)$.  \edit{In this paper we will assume that all galaxies
are viewed through a common lensing \vecg, and that the prior
$P(\vecg)$ is much less informative than the data and can be taken as
uniform.  Thus we focus on determining $P(\vecD | \vecg)$.  We leave
to future work the extension of the method to other circumstances,
such as when \vecg\ is known to follow some parametric form of
position (discussed in BA14), or when \vecg\ is drawn from a Gaussian
random field.}

\subsection{Simplest case}
Ultimately we will need to determine $P(\vecD | \vecg)$ in the case
where the data contain images of an arbitrary number of galaxies at
unknown locations.  We will assume that the pre-seeing, pre-lensing
images are drawn from a known library of ``template'' galaxies,
indexed by $G$, which in practice we will obtain by observing a
fraction of our survey to significantly higher $S/N.$
We begin, however, with a simple case, in which we
know we are observing a single galaxy known to have underlying
template index $G$.  The
position on the sky of some reference 
point in the galaxy (such as its centroid) we denote as $\vecx_G.$
Knowing $G$ we simulate the action of the lensing distortions and the
observing process (namely the PSF and pixelization)
to predict the data vector $\vecD^G(\vecg,\vecx_G)$ that we would obtain
from a noiseless observation.  The observed data vector is
\begin{equation}
\vecD = \vecD^n + \vecD^G(\vecg, \vecx_G),
\label{sumdata}
\end{equation}
and we assume that we know the likelihood function $\likeli(\vecD^n)$
of the added noise.
In the case that $\vecx_G$ is known, we have
\begin{equation}
P(\vecD | \vecg) = \likeli(\vecD^n) = \likeli\left[\vecD - \vecD^G(\vecg,\vecx_G)\right].
\end{equation}

A central strategy of BFD is to compress the pixel data to a short
vector \vecM\ 
that carries most of the information about lensing distortion.  The
critical requirement on the compression is that we are able to
propagate the distribution of $\vecD^n$ into a probability $P(\vecM |
\vecM^G)$ of observing compressed data \vecM\ given that the noiseless
underlying galaxy image compresses to $\vecM^G$.  This is most
straightforwardly accomplished by having the compression be a linear
operation on \vecD\ such that \eqq{sumdata} becomes
\begin{equation}
\vecM = \vecM^n + \vecM^G(\vecg, \vecx_G),
\label{summoments}
\end{equation}
and we will have, for fixed $\vecx_G,$
\begin{equation}
 P(\vecM | \vecg) = \likeli(\vecM^n) = \likeli\left[\vecM - \vecM^G(\vecg,\vecx_G)\right].
\end{equation}

We choose for \vecM\ a set of moments of the Fourier transform $\tilde
I^o(\veck)$ of the observed surface brightness $I^o(\vecx),$ defined
as
\begin{equation}
\label{ft}
\tilde I^o(\veck;\vecx_0) \equiv \int d^2x \, I^o(\vecx)
e^{-i\veck\cdot(\vecx-\vecx_0)}
\end{equation}
\edit{Note that this compression requires a choice $\vecx_0$ of coordinate
origin.  In this section we will assume that we have \textit{a priori}
knowledge of $\vecx_G$ and can set $\vecx_0=\vecx_G.$  In the next
section we will develop a treatment for the case of unknown $\vecx_G.$}
The data
\vecD\  are a regular sampling of $I^o,$ so in practice the Fourier
transforms are discrete.  We choose the compressed data vector
\begin{equation}
\label{moments}
\vecM(\vecx_0) \equiv \left( \begin{array}{c}
M_f \\
M_r \\
M_+ \\
M_\times
\end{array}
\right) = \int d^2k\, \frac{\tilde I^o(\veck;\vecx_0)}{\tilde T(\veck)}
W(|\veck^2|) \vecF; \qquad \vecF \equiv
\left( \begin{array}{c}
1 \\
k_x^2 + k_y^2 \\
k_x^2 - k_y^2 \\
2 k_x k_y
\end{array}
\right).
\end{equation}
where $\tilde T(\veck)$ is the Fourier transform of the PSF that has
convolved the observed image.  $W(|\veck^2|)$ 
is a real-valued window function applied to the integral to bound the noise, in
particular confining the integral to the finite region of $\veck$ in which
$\tilde T(\veck)$  is non-zero.  We calculate the moments in Fourier
domain in order to simplify the removal of the effects of the PSF, but
these moments are equivalent to taking radially weighted zeroth and
second moments of the real-space, pre-seeing image of the galaxy.

The moments are \emph{not} normalized, so that 
$\vecM$ remains a linear function of $\vecD$. The noise moment vector,
being a sum over the statistically independent noise of many pixels,
will have a likelihood $\likeli(\vecM^n)$ rapidly tend toward a multivariate Gaussian 
with covariance matrix $\matC_M.$  We assume that \emph{the pixel
  noise $\vecD^n$ is stationary,} in which case there is no covariance
between the noise at distinct \veck\ values, and the covariance matrix
elements are related to the power spectrum $P_n(\veck)$ of the noise by
\begin{equation}
\left(\matC_M\right)_{ij} = 
\int d^2k\, P_n(\veck) \left| \frac{W(|\veck^2|)}{\tilde T(\veck)}
\right|^2 F_i(\veck) F^\star_j(\veck).
\label{Cm}
\end{equation}
Note that while background shot noise and detector
read noise are stationary, any significant shot noise from the
galaxy's photons will violate stationarity.  With sensible choice of
$W$, the moments \vecM\ carry most of the information available about
shear of the source \citep{bj02}.  There are many practical benefits to
discarding the rest of the information in \vecD, as will become
apparent, but we highlight first that $\vecM^G$ is
independent of the observational conditions, \ie\ has been corrected
for the PSF, so we do not need to recalculate $\vecM^G$ as the PSF
varies, as long as we hold $W$ fixed.

\edit{We note at this point that there is much freedom in the choice
  of $W$.  As long as $W$ leads to finite noise $\matC_M$ via
  \eqq{Cm}, BFD remains valid and unbiased; but the accuracy of the
  inference on $\vecg$ will depend upon the choice of $W$.  One may
  choose to adjust this weight to optimize shear inference for a given
  set of observing conditions, but it is critical that the choice does
  \emph{not} depend upon the properties of target galaxies.  In this
  sense the BFD method is sub-optimal, in that it may not
  extract the most precise measure of \vecg\ from both
  large and small galaxies simultaneously.  We
  describe one choice for $W$ in Section~\ref{weightsec}.}

The next assumption in BFD is that \emph{the lensing is weak, so that
  a second-order Taylor expansion about $\vecg=0$ fully describes
  $P(\vecM | \vecg)$ for observed values of \vecg. }  In this case we have
\begin{align}
P(\vecM | \vecg) & = P + \vecQ\cdot \vecg + \frac{1}{2} \vecg \cdot
\matR \cdot \vecg, \\
P & \equiv P(\vecM | \vecg=0)  = \likeli\left(\vecM -
  \vecM^G\right)
 = |2\pi\matC_M|^{-1/2} \exp\left[-(\vecM - \vecM^G)^T
  \matC_M^{-1}(\vecM - \vecM^G) \right], \\
\label{q1eqn}
\vecQ & \equiv \left.\bnabg P(\vecM | \vecg)\right|_{\vecg=0}=
        -\bnabg \vecM^G \cdot \bnab_{M} \likeli\left(\vecM-\vecM^G\right) \\
\matR & \equiv \left. \bnabg \bnabg P(\vecM | \vecg)\right|_{\vecg=0}. 
\end{align}
At the end of \eqq{q1eqn} we have assumed
that \emph{the noise
  likelihood is invariant under shear of the underlying galaxy $G$} so
that we can propagate all shear derivatives into derivatives of the
properties of the template galaxy.
This is satisfied for background-limited images.
The quantities \vecQ\ and \matR\ give the differential probability of
observing the image under lensing distortions.  If \vecD\ is comprised
of many independent
observations $\vecD_i$ of the same underlying galaxy $G$ with the same applied
lensing, we can produce the quantities $P_i, \vecQ_i, \matR_i$ as above for
each observation, then the total posterior probability for \vecg\ is
given by
\begin{align}
-\ln P(\vecg | \vecD) & = (\rm const) - \ln P(\vecg) - \sum_i \ln
                        P(\vecD_i | \vecg) \\
\label{pqr1}
 & =  (\rm const) - \ln P(\vecg) - \vecg \cdot \vecQ_{\rm tot}
+ \frac{1}{2} \vecg \cdot \matR_{\rm tot}
   \cdot \vecg, \\
\label{qtot}
\vecQ_{\rm tot} & \equiv \sum_i   \frac{\vecQ_i}{P_i} \\
\label{rtot}
\matR_{\rm tot} & \equiv \sum_i \left(
   \frac{\vecQ_i\vecQ_i^T}{P_i^2} - \frac{\matR_i}{P_i}\right)
\end{align}
The posterior distribution is, ignoring the prior, Gaussian in \vecg,
with inverse covariance matrix
\begin{equation}
\label{cg}
\matC_g = \matR_{\rm tot}^{-1}
\end{equation}
and mean value
\begin{equation}
\label{gbar}
\bar\vecg = \matR_{\rm tot}^{-1} \vecQ_{\rm tot}.
\end{equation}

\subsection{Detection and selection}
Consider now the case where there is a galaxy present but we do not
know its position $\vecx_G$ in advance, \edit{so we need a
  prescription for choosing those locations $\vecx_0$ about which we
  will compute moments.} We need a \emph{detection}
process to decide if the galaxy has been observed within some small
region $\Delta^2x$ about some position $\vecx_0$.  Once a detection is
made, we will also require some \emph{selection} criteria to decide
which detections will be used to constrain the lensing.  \edit{In this section, we will continue to assume that the
unlensed appearance $G$ of the galaxy is known, but its location is
not.} At each
potential source location, we end up with either a
successful detection and selection, plus measured moments $\vecM$; or
a non-selection.  We therefore need to know $P(\vecM,s,d|G) = P(\vecM,s | G)$ for the
former case, and $1-P(s|G)$ for the latter case, where $s$ ($d$)
indicates successful selection (detection).

These probabilities are readily calculable if we make the detection
and selection using the compressed quantities themselves.  We add to
our compressed data set the two weighted first moments of the source
in Fourier space:
\begin{equation}
\vecX(\vecx_0) \equiv \int d^2k\, \frac{\tilde I^o(\veck; \vecx_0)}{\tilde T(\veck)}
W(|\veck^2|) 
\left( \begin{array}{c}
ik_x \\
ik_y
\end{array}
\right).
\end{equation}
We choose as a criterion for detection of a source at $\vecx_0$ that
$\vecX(\vecx_0)=0$.  Our choice of moments for $\vecM$ and $\vecX$ have these
useful properties:
\begin{align}
\frac{dM_f}{d\vecx_0} & = \vecX, \\
{\rm Cov}(\vecM, \vecX) & = 0 \quad \mbox{(for stationary noise)}, \\
\label{Jdef}
J \equiv \left| \frac{d \vecX}{d \vecx_0} \right| & =
                                                          \left(M_r^2
                                                          - M_+^2 -
                                                          M_\times^2\right)/4
                                                                  =
                                                                  \vecM^T
                                                                  \matB
                                                                  \vecM,
  \\
\matB & \equiv  {\bf diag}(0,1/4,-1/4,-1/4), \\
\langle J^n \rangle & = {\rm Tr}(\matB\matC_M)=0, 
\end{align}

The first line
means that we detect a source at all stationary points of the function
$f(\vecx_0)=M_f$, the zeroth moment of the image as convolved with a
filter defined by $W(|\veck^2|)/\tilde T(\veck).$  This
filter will be broader than the PSF in any sensible application of
BFD. The second property yields $\likeli(\vecM^n,\vecX^n) =
\likeli(\vecM^n)\likeli(\vecX^n)$ for our multivariate Gaussian noise
distribution.  The third property shows that the Jacobian
determinant $J$ of the positional moments \vecX\ is purely a function
of $\vecM$, and hence statistically
independent of \vecX.

\edit{
The noiseless moments expected from galaxy $G$ are now functions 
$\vecD^G(\vecg, \vecu=\vecx_G-\vecx_0)$ and
$\vecX^G(\vecg, \vecu)$ since it is only the displacement \vecu\
between the galaxy location and the Fourier phase center that
matters.  The detection condition is 
$\vecX=\vecX^G+\vecX^n=0$ so the probability of this occurring in a 
small region $\Delta^2x$ about $\vecx_0$ is 
\begin{align}
P(\vecM,d | G, \vecg, \vecx_G, \vecx_0) & =
\likeli\left[\vecM-\vecM^G(\vecg,\vecx_G-\vecx_0)\right] 
          \likeli\left[-\vecX^G(\vecg,\vecx_G-\vecx_0)\right]
          \left| \frac{d\vecX}{d\vecx_0}\right| P(\vecx_0) \, \Delta^2x \\
 & =
\likeli\left[\vecM-\vecM^G(\vecg,\vecx_G-\vecx_0)\right] 
          \likeli\left[\vecX^G(\vecg,\vecx_G-\vecx_0)\right]
          \left| J \right| \, \Delta^2x. 
\end{align}
In the last line, we take advantage of (\ref{Jdef}), assume a uniform
prior $P(\vecx_0)$ on the position of the detection, and note that
the zero-mean, multivariate Gaussian will have $\likeli(-\vecX^G)=\likeli(\vecX^G).$}

To eliminate noise detections, we will want to discard low-flux
detections.  We implement the selection criterion as membership in a
subregion \edit{$S$} of moment space:
\begin{align}
S: & f_{\rm min} < f < f_{\rm max} \\
\label{pMsG1}
\Rightarrow \quad P(\vecM,s | G,\vecg, \vecx_G, \vecx_0) & = \left\{
\begin{array}{ll}
 \likeli(\vecM-\vecM^G) \likeli(\vecX^G) |J| \Delta^2 x& \vecM \in S \\
 0 & \vecM \notin S
\end{array} \right. \\
\label{psG1}
\Rightarrow \quad
P(s | G, \vecg, \vecu=\vecx_G-\vecx_0) & = \likeli(\vecX^G) \Delta^2x \int_{\vecM\in S} d\vecM\,
           \likeli(\vecM-\vecM^G) |J(\vecM)|.
\end{align}
\edit{For brevity we suppress the dependence of the template galaxy's moments
$\vecM^G,\vecX^G$ on the applied lensing \vecg\ and on the
displacement $\vecx_G-\vecx_0$ between the galaxy position and the
detection location.}

To render the integration in (\ref{psG1}) tractable, we make the
simplifying assumption that \emph{the Jacobian determinant $J$ of the
  first moments is positive at any location where there is
  non-negligible probability of selection:}
\begin{equation}
J = \vecM^T \matB \vecM = J^G + 2\left(\vecM^G\right)^T \matB \vecM^n
+ J^n > 0.
\end{equation}
Since $J$ is the determinant of the 2nd derivative matrix of $f$, a
restatement is that we are assuming the $f(\vecx_0)$ surface is (nearly) always convex if $f_{\rm
  min}<f<f_{\rm max}.$  To maintain this approximation we will need to avoid
noise detections by raising $f_{\rm min}\gtrsim 5\sigma_f,$ where we
define
\begin{equation}
\sigma_f^2 = \left(\matC_M\right)_{ff}.
\label{sigmaf}
\end{equation}

We discuss this convex-detection approximation in Section~\ref{convexsec}.
With this approximation, we can integrate a multivariate Gaussian
$\likeli$ in \eqq{psG1} analytically, obtaining
\begin{align}
\label{psG2}
P(s | G, \vecg, \vecu ) & = \likeli(\vecX^G) \Delta^2 x \left[
 J^G Y + 2\left(\matC_M \matB \vecM^G\right)_f \frac{\partial
  Y}{\partial f_G}
+ \left(\matC_M \matB \matC_M\right)_{ff} 
\frac{\partial^2
  Y}{\partial f_G^2} \right], \\
\label{gaussY}
Y & \equiv (2\pi)^{-1/2} \int_{(f_{\rm min}-f_G)/\sigma_f}^{(f_{\rm
    max}-f_G)/\sigma_f} d\nu\, e^{-\nu^2/2}.
\end{align}

Now consider the joint distribution of the detection/selection
outcomes at a grid $\vecx_1, \vecx_2, \ldots,
\vecx_j \ldots$ of all search positions with non-negligible
selection probability $P(s|G, \vecx_G, \vecx_0=\vecx_j)$.
We assume now that \emph{galaxies are uncrowded, in that
  no other galaxies contribute significantly to $\vecM$ or $\vecX$ at
  any location $\vecx_j$ where galaxy $G$ might be selected.}  At each
search position, we either have a selection and a resultant \vecM, or
we have a non-selection.  If the search region is contiguous, there
can be \emph{at most one} of the
$\vecx_j$ with successful selection.  This follows from our assumption
that $J>0$, which implies that that map $\vecx_0\rightarrow\vecX$ is
one-to-one over a contiguous region, so that 
$\vecX=0$ can only occur at a single $\vecx_0.$


With this single-selection rule, we have two possible outcomes:
\begin{enumerate}
\item A detection at a single location $\vecx_j$ yielding moments \vecM,
  with probability $P(\vecM, s_j|G)$ from \eqq{pMsG1}, or
\item No detection at all, with probability $1-\sum_j P(s_j | G),$
  using the selection probability in \eqq{psG2}.
\end{enumerate}
Integrating over all possible detection positions, we obtain a total
probability of outcome (1):
\begin{align}
P(\vecM, s | G, \vecg, \vecx_G) & =  J(\vecM) \int d^2u\, \likeli\left[\vecX^G(\vecg,\vecu)\right] 
\likeli\left[\vecM-\vecM^G(\vecg,\vecu)\right]. \\
\label{pMsG2}
& \approx  J(\vecM) \sum_{\vecu} \Delta^2u\, \likeli\left[\vecX^G(\vecg,\vecu)\right] 
\likeli\left[\vecM-\vecM^G(\vecg,\vecu)\right].
\end{align}
In the second line, we change the
integration to a sum over a 2d grid of points $\vecu$ with cell area
$\Delta^2u,$ since this is how we implement the integration over
source position. We can
truncate the grid where $P(s_j|G)$ becomes negligible.  As expected,
the resulting probabilities are independent of both the true position
$\vecx_G$ of the galaxy and the position $\vecx_i$ of the detection
once the observed moments \vecM\ are specified.

The total probability of detection is obtained by similarly
integrating \eqq{psG2} over all \vecu:
\begin{equation}
\label{psG3}
P(s | G, \vecg) = \sum_{\vecu} \Delta^2u \, \likeli\left(\vecX^G\right) \left\{
 J\left(\vecM^G\right) Y + 2\left(\matC_M \matB \vecM^G\right)_f \frac{\partial
  Y}{\partial f_G}
+ \left(\matC_M \matB \matC_M\right)_{ff} 
\frac{\partial^2
  Y}{\partial f_G^2} \right\},
\end{equation}
remembering that $\vecM^G, \vecX^G, f_G,$ and the arguments to $Y$
depend upon \vecg\ and \vecu.  For
a galaxy with flux $f_G$ that is many $\sigma_f$ away from the
selection boundaries, we have $Y\rightarrow 1$.  In this case it is
easy to see that $P(s | G,\vecg)\rightarrow 1$, by recasting (\ref{psG3}) as
an integral over $\vecX^G$---as long as $J>0$.  If the positive-$J$
assumption does not hold, \eqq{psG3} is incorrect, and
we can have a mean number of detections per
source that is $>1$.  In Section \ref{convexsec} we discuss our approach to
mitigating failure of the positive-$J$ assumption.

\subsection{Galaxy populations: postage stamp case}
\label{stampsec}
Now we generalize from having a single galaxy type $G$ to having $G$
be an index into the entire catalog of possible galaxy images.
\edit{We assume we know the prior probability $p_G$ that a galaxy is
  of type $G$.  If for example the galaxy library is approximated by
  the set of galaxies found in a high-$S/N$ imaging survey of the sky,
  each detected galaxy would be assigned equal $p_G$.}

First consider the artificial case (commonly used in shear-testing
programs) in which we know that exactly one galaxy has been placed
in each of many disjoint ``postage stamps'' of pixels $\vecD_i \in
\vecD$.  In each stamp, we either obtain a selection with measurement
of moments $\vecM_i$ at some location in the stamp, or we obtain a
non-selection.  The probabilities of these two outcomes are
\begin{align}
\label{pMs1}
P(\vecM_i, s|\vecg) & = J(\vecM_i) \sum_{G,\vecu}  p_G\Delta^2u\, \likeli(\vecX^G) 
\likeli(\vecM_i-\vecM^G), \\
P(\sim s | \vecg) & = 1 - P(s|\vecg) \\
\label{ps1}
P(s|\vecg) & = \sum_{G,\vecu} p_G \Delta^2u\,\likeli(\vecX^G) \left[
 J(\vecM^G) Y + 2(\matC_M \matB \vecM^G)_f \frac{\partial
  Y}{\partial f_G}
+ \left(\matC_M \matB \matC_M\right)_{ff} 
\frac{\partial^2
  Y}{\partial f_G^2} \right]
\end{align}
These are the key equations for the BFD calculation.  We have made
implicit the dependence of the noiseless template moments $\vecM^G$
and $\vecX^G$ on the source position $\vecu$ and the lensing \vecg.
We define as before the Taylor expansions
\begin{align}
P(\vecM_i,s | \vecg) & \approx P_i + \vecQ_i\cdot \vecg +
  \frac{1}{2}\vecg\cdot\matR_i\cdot\vecg, \\
P(s | \vecg) & \approx P_s + \vecQ_s\cdot \vecg +
  \frac{1}{2}\vecg\cdot\matR_s\cdot\vecg,
\end{align}
where $\vecQ_i=\bnabg P(\vecM_i,s),$ etc., are derived by propagating
derivatives through to template quantities $\vecM^G$ and $\vecX^G$.
\edit{The detection
probability $P(s)$ is integrated over all possible selected
moments and all possible galaxies $G$, so it does not depend on the data in stamp $i,$
only on the noise level and PSF of the observation as manifested in 
the covariance matrix $\matC_M$ in each stamp.}  For
notational simplicity we will assume here that all stamps have the
same noise level and PSF and hence the same $\matC_M,$ but the
formalism and our 
implementation allow for variation between targets.

The combined probability of the output of the
observation/detection/selection/compression process is
\begin{equation}
P(\vecD | \vecg) = P(\sim s | \vecg)^{N_{ns}} \prod_{i\in{\rm
    selections}} P(\vecM_i,s | \vecg) 
\end{equation}
where $N_{ns}$ is the number of non-selected stamps.  We can now
calculate the probability of the lensing variables, following \eqq{pqr1}:
\begin{align}
\label{pqr2}
-\ln P(\vecg | \vecD) & = (\rm const) - \ln P(\vecg) - \vecg \cdot \vecQ_{\rm tot}
+ \frac{1}{2} \vecg \cdot \matR_{\rm tot}
   \cdot \vecg, \\
\label{qtot2}
\vecQ_{\rm tot} & \equiv \sum_i   \frac{\vecQ_i}{P_i}  -
                  N_{ns}\frac{\vecQ_s}{1-P_s} \\
\label{rtot2}
\matR_{\rm tot} & \equiv \sum_i \left(
   \frac{\vecQ_i\vecQ_i^T}{P_i^2} - \frac{\matR_i}{P_i}\right) +
                  N_{ns}\left( \frac{\vecQ_s\vecQ_s^T}{(1-P_s)^2} + \frac{\matR_s}{1-P_s}\right)
\end{align}

We now have all the tools needed to make a lensing inference from a
postage-stamp data set.  We assume that we have available a
\emph{complete catalog of possible galaxies $G$} and that for each we
have a \emph{noiseless, unlensed image.}  In practice
of course our template set will be a finite sample from the (infinite)
distribution of detectable galaxies.  It is essential that the template
set is a fair sample of all galaxy types that can meet the selection
criteria with non-negligible probability.
In other words we must know about galaxies that are outside
the flux selection cuts by up to several $\sigma_f$.

The input data are: postage stamps of the
``observed'' galaxies, which we call the \emph{targets;}
low-noise postage stamp images of unlensed \emph{template}
galaxies to serve as our sample $G$; the PSF for each stamp; and
the noise power spectrum $P_n$ for each stamp.  Our testing assumes
white noise, $P_n=n.$

The procedure is as follows:
\begin{enumerate}
\item Select a weight function $W$ that will be applied to all targets
  and templates. The best choice will usually be a rotationally
  symmetric approximation to $\tilde T(\veck)^2 \tilde I_g(\veck)$, where
  $\tilde I_g$ is the transform of the unlensed, pre-seeing image of a
  galaxy of typical size in the survey.
\item For each template galaxy $G$, measure the moments $\vecX^G$ and
  $\vecM^G$ under $W$ for copies of the galaxy translated over a grid
  of $\vecx_G$ centered on the primary flux peak.  
  We can purge from the template set any that have negligible
  $P(s|G).$  Further calculate the first and second derivatives of all
  moments with respect to \vecg, using the formulae in Appendix~\ref{momentcalcs}.
\item For each target galaxy:
\begin{enumerate}
\item Find the point(s) near the object centroid where the detection
  criterion $\vecX=0$ is met.
\item Calculate the moments $\vecM_i$ about the detection point(s) and
  discard those failing the selection cut on the flux moment.  After
  this step we require \emph{no further access to the image data.}
\item If no selection is made, increment the count $N_{ns}$ of
  non-selections, and continue with the next stamp.  If more than one
  selection is made, choose the brightest and note that we have
  violated one of our assumptions!
\item Calculate $\matC_M$ for this stamp.
\item For each target postage stamp $i$, calculate $P_i =
  P(\vecM_i,s|\vecg=0)$ from \eqq{pMs1}, and also the derivatives under
  lensing $\vecQ_i$ and $\matR_i$.  Since this operation
  is executed for every target-template pair, it is the computational
  bottleneck of the procedure.  The summand in (\ref{pMs1})
  is simple, involving some 4-dimensional matrix
  algebra and one exponential, so is far faster than an
  iteration of a forward-modeling procedure.  The $\{P_i,\vecQ_i,\matR_i\}$ data
  fully encapsulate the lensing information from this galaxy and go
  into our catalog.
\end{enumerate}
\item Calculate the selection probability $P(s|\vecg=0)$ from \eqq{ps1},
  and its derivatives $\vecQ_s,\matR_s$ with respect to lensing.  Note
  this needs to be done only once for each distinct $\matC_M$.
\item Sum the contributions to the posterior $-\ln P(\vecg | \vecD)$ from detections and
  non-detections as given in
  Equations~(\ref{qtot2}) and (\ref{rtot2}).
\label{nsstep}
\item Add the Taylor expansion of any prior $P(\vecg)$ to $\vecQ_{\rm tot}$ and $\matR_{\rm tot}.$
\item We now have the posterior log probability for \vecg.  The shear
  estimate and its variance are in Equations~(\ref{gbar}) and
  (\ref{cg}).
\end{enumerate}

\subsection{Poisson-distributed galaxies}
For real sky images, we replace the postage-stamp distribution of
galaxies with a Poisson distribution.  We assume a total unlensed
density $n$ of sources on the sky, with probabilities $p_G$ of each
galaxy being of type $G$.
If our target survey spans solid angle
$\Omega$ of sky, consider dividing this area up into regions of area
$\Delta\Omega$ larger than the selection region of any single galaxy,
but small enough that $n\,\Delta\Omega\ll1$ so that we only
have 0 or 1 galaxy in the region after running the
detection/selection/compression process across the survey.  The
probability of obtaining a detection with moments $\vecM_i$ within any
small sky area $\Delta\Omega$ is
\begin{align}
P(\vecM_i, s | \vecg, \Delta\Omega) & = \sum_G P(\vecM_i, s | \vecg,
                                      G) P(G | \Delta\Omega) \\ 
 & = n\,\Delta\Omega P(\vecM_i, s | \vecg),
\end{align}
where we take $P(\vecM_i,s | \vecg)$ from \eqq{pMs1}.  Similarly, the
probability of selecting a source in a single cell 
\begin{align}
P(s | \vecg, \Delta\Omega) & = n\,\Delta\Omega \sum_G p_G P(s | G, \vecg), \\
  & = n\, \Delta\Omega\, P(s|\vecg),
\end{align}
where we use $P(s|\vecg)$ from \eqq{ps1}.
The quantity $nP(s|\vecg)$ is the expected sky density
of selected galaxies.  It depends on \vecg\ through the moments of the
template galaxies, as per usual.

Our total
data \vecD\ are reduced to a list $\{\vecM_i, \vecx_i\}$ for $1\le i
\le N_s$ of the locations and moments of the $N_s$ selected sources;
plus the information that there are no selections at any other
locations.  The total posterior for \vecg\ is now
\begin{align}
P(\vecg | \vecD) & \propto P(\vecg) 
\prod_{\rm non-detections} \left[1-P(s | \vecg, \Delta\Omega)\right] \quad
                   \prod_{i=1}^{N_s} P(\vecM_i, s | \vecg, \Delta\Omega)\\
\label{poissonposterior}
 & = P(\vecg)  e^{-n\Omega P(s|\vecg)} (n\Delta\Omega)^{N_s} \prod_{i=1}^{N_s} P(\vecM_i,
   s|\vecg).
\end{align}
The $(\Delta\Omega)^{N_s}$ term is independent of \vecg\ and can be
dropped.  We retain dependence on $n$ since we may wish to consider
the source density as a free parameter along with \vecg\ if we are
simultaneously constraining source clustering and shear.
This posterior differs from the postage-stamp case only in
the non-selection term.  We replace (\ref{qtot2}) and (\ref{rtot2}) with
\begin{align}
\label{pqrpoisson}
-\ln P(\vecg | \vecD) & = ({\rm const}) - \ln P(\vecg) - N_s\log n +
                        n\Omega P_s - \vecg \cdot \vecQ_{\rm tot}
+ \frac{1}{2} \vecg \cdot \matR_{\rm tot}
   \cdot \vecg, \\
\label{qtotpoisson}
\vecQ_{\rm tot} & \equiv \sum_i   \frac{\vecQ_i}{P_i}  - n\Omega
                  \vecQ_s, \\
\label{rtotpoisson}
\matR_{\rm tot} & \equiv \sum_i \left(
   \frac{\vecQ_i\vecQ_i^T}{P_i^2} - \frac{\matR_i}{P_i}\right) +
                  n\Omega \matR_s
\end{align}
 
The operative procedure for inferring shear from a sky image is hence
identical to that given for the postage-stamp case, except that of
course we search the entire image for detections, not just the centers
of each stamp.  We use the above formulae in step~\ref{nsstep} instead
of the postage-stamp formulae.

\subsection{Sampling the template space}
\label{templates}
The BFD method depends upon approximating the full galaxy population
with a finite sample of galaxies $G$ from the sky.  In essence we are
approximating the continuous distribution of galaxies in the moment
space with a set of $N_G$ $\delta$ functions at a random sampling from
the distribution.  The measurement error distribution
$\likeli(\vecM-\vecM^G)$ acts as a smoothing kernel over the samples.
 While the sums over $G$ for $P_i$ (and $\vecQ_i, \matR_i$) in \eqq{pMs1}
are unbiased estimates of the complete integrals over moment space,
there are two issues we must address.  

First, in producing $\vecQ_{\rm tot}$ and $\matR_{\rm tot}$ we divide
$\vecQ_i$ and $\matR_i$ by $P_i$.  As noted in BA14, division by a
noisy estimator for $P_i$ produces a bias that scales inversely with the number
of template galaxies contributing significantly to the $P_i$ sums.
The number of galaxies we can measure at sufficiently high $S/N$ to
use as templates will be limited by scarce observing time.
Fortunately we can increase the density of templates in
moment space by exploiting the rotation and parity symmetry of the
unlensed sky: for each $G$ that we observe, we can assume that rotated
and reflected copies of this galaxy are also equally likely to
exist. In practice we partition $p_G$ among such copies and add them
to the template set. We will investigate in Section~\ref{samplesec} the bias
resulting from finite template sampling.

Second: 
because our \vecM\ consists of un-normalized moments, the spacing
between template galaxies in moment space will become large compared
to the measurement error ellipsoid described by $\matC_M$ when we
observe target galaxies at high $S/N$.  Bright targets can easily end
up with no templates for which $\likeli(\vecM - \vecM^G)$ is
non-negligible.  Even worse, the $P_i$ sum for a galaxy can be
dominated by a single template that is many $\sigma$ away from the
target in moment space, and this produces large derivatives in $\ln
P(\vecM_i,s)$ with respect to \vecg, giving spuriously large influence in the
final lensing estimator. It is further true that brighter galaxies are rarer
on the sky, so our template survey will contain fewer sources with
flux comparable to our brighter targets.

It is therefore advantageous to \emph{add noise to the moments
  measured for bright galaxies.}  One may question the sanity of
adding noise to hard-won signal, but note that weak shear (magnification)
measurements accrue uncertainty from the intrinsic variation of galaxy shapes
(sizes) as well as from the measurement noise in these quantities.
Typically, once
$S/N\gtrsim 20$, the intrinsic variation of the population is the
dominant form of noise.  So a resolved galaxy with $S/N\approx75$
loses little lensing information
if degraded to $S/N\approx 25$. 
However if we triple the
noise, the likelihood function will
``touch'' $3^4 \times$ more template galaxies in our 4-dimensional
$\vecM$ space, so we can reduce template sample variance and bias by
increasing noise.  

We must be careful to implement this process such that
$P(\vecM, s | G, \vecg)$ remains calculable for both the bright
galaxies and faint ones.  Again this is best done by using the moments
themselves to decide whether to add additional noise.  The procedure
that we use is as follows; in Appendix~\ref{addnoise} we present the
altered formulae for $P(\vecM, s |G, \vecg)$ that apply to the galaxies which
have had noise added.
\begin{enumerate}
\item We establish bounds $f_1$ and $f_2$ on the galaxies to which we
  wish to add noise, based on comparing
  the density of templates with the covariance matrix $\matC_M$ of 
the measured moments.
\item We detect, measure, and select target galaxies the same way as
  described in Section~\ref{stampsec}, in the flux range $f_1<f<f_2.$
\item For each selected galaxy, we form a new moment vector ${\mathcal
    M}=\vecM + \vecM_A$, with $\vecM_A$ drawn 
  from a multivariate Gaussian with zero mean and predetermined
  covariance matrix $\matC_A$.  
  We make no further use of the original moments $\vecM$.
\item We proceed with the analysis as before, with the exception that
  $P({\mathcal M}, s | G, \vecg)$ from \eqq{addpMs} is used in place of our
  previous $P(\vecM,s | G, \vecg)$.  
  Note the probability $P(s|\vecg)$ of galaxy selection in
  \eqq{ps1} remains accurate, since selection is made before adding
  noise to the moments.
\end{enumerate}

More generally we may define a series of $b$ flux bins by bounds
$f_0,f_1,\ldots,f_b$, and choose for each bin a distinct
covariance matrix $\matC_A$ for the added noise (presumably adding
zero noise in the lowest-flux bin).  For each target galaxy we calculate
$P({\mathcal M},s | G, \vecg)$ using the value of $\matC_A$ we have applied.
The non-selection term $P(s|\vecg)$ is calculated using $f_{\rm min}=f_0,$
$f_{\rm max}=f_b.$  The only requirement on the added noise is that it
obey the condition ${\rm Tr}(\matB\matC_A)=0$ which holds for
stationary noise.

\section{Implementation}
\label{algorithm}
We have implemented the BFD shear inference in C++ code.
The computational bottleneck of the BFD method is the evaluation of
$P(\vecM,s|G, \vecg)$, which must be done for each target-template pair.  A
survey like \emph{DES} might detect $\sim10^{8.5}$ galaxies, and use
$\sim10^{4.5}$ templates, each replicated over $\sim10^4$ different
translations and rotations, leading to $\sim10^{17}$ evaluations of
$P(\vecM,s | G, \vecg)$.

Substantial speedup is attained if we can rapidly cull the templates
to those which make significant contributions to the sums for $P_i,
\vecQ_i,$ and $\matR_i$, \ie\ eliminate those highly suppressed by the
Gaussian exponential in \eqq{pMs1}. In this Section we describe some
shortcuts to reduce the scale of the problem, and
an efficient algorithm for culling the target-template pairs, which
leads to an implementation that is feasible to run on modest
present-day hardware for even the largest foreseen surveys.

\subsection{Computational shortcuts}
The target galaxies all have $\vecX=0$ by definition of the detection
criterion, and so we may first eliminate any template with
small $\likeli(\vecX^G),$ a criterion we use to bound the
displacements $\vecu$ at which we replicate the templates.
Furthermore we have the freedom to rotate the coordinate axes for
each target by the
angle $\beta$ which sets one of the ellipticity moments $M_\times=0$.
We must rotate $\matC_M$ into this frame, and make sure to rotate all
the $\vecQ_i$ and $\matR_i$ back to the original coordinate system after
each is calculated.  The unlensed population must be invariant under coordinate
rotation, so we do \emph{not} have to rotate the $\vecM^G.$
With this procedure, we can prune the
templates to those that are within $\sim 6\sigma$ of $M_\times=0.$
The space ${\vecM^G,\vecX^G}$ of template moments is
now bounded to a small interval near the origin in 3 of its 6 dimensions.

\subsection{$k$-d tree algorithm}
\label{kdtree}
In building the prior we need to efficiently identify template
galaxies with moments $\vecM^G$ that
are close, in moment space, to a given target
galaxy \vecM.  The relevant equation is
\begin{equation}
\label{chisq}
\chi^2 \equiv \left(\vecM-\vecM^G\right)^T \matC_M^{-1}
\left(\vecM-\vecM^G\right) \le \sigma_{\rm max}^2.
\end{equation}
We must be careful in choosing $\sigma_{\rm max}$ so that 
truncation of the integral does not bias \vecg; but the number of
sampled template galaxies, and the execution time of the measurement,
will scale as $\sigma_{\rm max}^6.$

We
choose to store the moments of the template galaxies in a $k$-d tree \citep{kdtree}, which
partitions the templates into distinct $k$-dimensional rectangular nodes that allow for
fast lookup of points satisfying (\ref{chisq}).  The $k$-d tree is
built by assuming a nominal covariance matrix $\matC_N$ that is close
enough to the $\matC_M$ of the targets that the set of templates satisfying
(\ref{chisq}) with $\matC_N$ includes all those which do for
$\matC_M,$ and not many more.
To reduce the number of computations, we do a Cholesky decomposition 
$\matC_N^{-1}=\matA^T\matA$, and
rescale the template and target moments to $\vecN\equiv \matA\vecM,
\vecN^G\equiv \matA\vecN^G$.  This transformation yields
$\chi^2 = |\vecN-\vecN^G|^2,$ the Euclidean distance in \vecN.    The
\vecN\ are used only to isolate the relevant templates, not to
calculate the probabilities.

We need to replicate each template at
a grid in $\vecu$ and rotation angle.  The step sizes in translation
and rotation are chosen such that $\vecN^G$ shifts by $\approx
\sigma_{\rm step}\lesssim 1$ between each grid point. Parity-reversed
copies are also made. The probability
$p_G$ of each template is shared equally between its copies.
We discard template copies
that have no chance of satisfying \eqq{chisq} for any
selected target galaxy (remembering that all selected targets have $\vecX=0,$
$M_\times=0,$ and $f_{\rm min}<M_f<f_{\rm max}$).

The derivatives of $\vecM^G$ with respect to shear are calculated for
all retained templates.  
If all the target galaxies have the same covariance matrix,  a number of numerical factors
can be precomputed so that they do not need to be recalculated for
every template/target pair.  Note that a new template set 
needs to be constructed, and the $k$-d tree partition repeated, if the
target $\matC_M$ changes by more
than $\approx10\%.$  The construction of the template tree scales as
$N_G\log N_G,$ where $G$ is the number of templates, which is subdominant to
the time $N_t N_G$ for integrating the $N_t$ targets over the template set.

After the tree has been constructed, we find for each target
galaxy all the nodes that contain template galaxies with
$\chi^2 < \sigma_{\rm max}^2$ using the nominal $\matC_N$.  
If the number of templates in the retained nodes exceeds $3N_{\rm
  sample},$ we randomly subsample a fixed number $N_{\rm sample}$ of
them according to their probabilities $p_G$.  This keeps us from
wasting time calculating huge numbers of template/target pairs for
targets with large uncertainties, while making full use of the
templates that resemble the rarer targets. 
With this list of template/target pairs, we can calculate the $P, \vecQ,$ and 
$\matR$ values needed.   The speed of the
integration step now scales as $N_tN_{\rm sample}$ \edit{if the number
  of templates becomes large.}

Our implementation executes the
integration over templates for $\approx10$ galaxies per second per
core on a general-purpose cluster, for the \galsim\ simulations below
in which each target is compared to $\approx40,000$ templates.  At
this speed, a 1000-core cluster 
could measure $10^9$ target galaxies (\eg\ the LSST survey) in just 1
day, probably much faster than the subsequent cosmological inferences
will require.

While the BFD method has no parameters to tune to reduce bias, the sampling/integration
algorithm has three free parameters---$\sigma_{\rm max}, \sigma_{\rm
  step},$ and $N_{\rm sample}$---which trade computational speed
and memory requirements against the bias induced by finite sampling.
The number $N_G$ of templates sampled from the sky also will be
important in controlling finite-sample biases.

\subsection{Weights and PSFs}
\label{weightsec}
The weight function $W(|k^2|)$ used in calculating the moments of
\eqq{moments} must satisfy two requirements: first, it must vanish at
any $\veck$ where $\tilde T(\veck)=0$, in order to keep measurement
errors finite; and it must have two continuous derivatives in order
for the shear derivatives of the template moments to be calculable
(see Appendix \ref{momentcalcs}).
With these conditions satisfied, BFD is well-defined and unbiased,
but further refinement of $W$ can optimize the noise on the inferred
\vecg\ and the required size of ``postage stamp'' of pixels for the DFT around
each galaxy.  In our validation tests we use this
``$k\sigma$'' weight function:
\begin{equation}
\label{ksigma}
W\left(|k^2|\right)  \equiv \left\{ 
\begin{array}{cc}
\left( 1 - \frac{k^2\sigma^2}{2N}\right)^N & k <
                                             \frac{\sqrt{2N}}{\sigma} \\
0 & k \ge
                                             \frac{\sqrt{2N}}{\sigma} 
\end{array}
\right.
\end{equation}
with $N=4.$  This closely approximates a Gaussian with width (in $k$
space) of $1/\sigma$, but goes smoothly to zero at
finite $k$.

In our validation tests we assume we have a noiseless, Nyquist-sampled
postage stamp of the PSF from which we can measure $\tilde T(\veck)$
on a discrete grid of $\veck$.  If we require $\tilde T$ at other
values of \veck, we interpolate the prescription for zero-padding in
real space and quintic polynomial interpolation in $k$-space given by 
\citet{kinterp}.  This need arises if there is distortion across the
image such that either targets or templates are sampled at slightly
different pitch than the PSF.

\section{Validation}
\label{validation}
To verify that our implementation of BFD can infer shear with an
accuracy of $|m|\lesssim10^{-3},$ we use two types of simulated data.  
The ``Gauss tests'' use Gaussian galaxies, a $\delta$-function PSF,
and a Gaussian $W(|k^2|),$ in which case we can calculate all moments
and their shear derivatives analytically---no rendering of images is
done, so this is fast and bypasses any issues related to image
discreteness.  The second validation test uses simulated galaxy
images produced with the Python/C$++$ software
\galsim\
\citep{galsim}.\footnote{https://github.com/GalSim-developers/GalSim} 

Table~\ref{sims} gives the parameters of the two validation
simulations.  While they use different methods to generate
``observed'' moments for the target and template galaxies, they use
the same integration code.  Both simulations proceed as follows:
\begin{enumerate}
\item A common galaxy generator is used to generate target and
  template samples, with shear and noise being applied only to the
  targets. The galaxies are sampled from a uniform
  distribution in $S/N$ (Gauss test) or flux (\galsim\ test) between
  specified limits.  The galaxy half-light radius $r_{50}$ is also
  drawn uniformly between two bounds.  The (unlensed) ellipticity
  $e=(a^2-b^2)/(a^2+b^2)$ of the source is drawn from the distribution
\begin{equation}
\label{eprior}
P(e) \propto e (1-e^{2})^{2} \exp\left(-e^{2}/2\sigma^{2}_e\right)
\end{equation}
and the galaxy position angle is distributed uniformly.  Galaxy
origins are randomized with respect to the pixel boundaries (if any).
\item A ``batch'' of measurements is made by generating $N_{\rm
    batch}$ target galaxies with a constant shear \vecg, adding noise,
  and measuring moments about the origin which yields $\vecX=0.$
  Those passing any selection cuts are integrated against $N_{\rm
    template}$ template galaxies drawn from the same generator, each
  of which is translated, rotated, and reflected as described above.
  The $P_{\rm tot}, \vecQ_{\rm tot},$ and $\matR_{\rm tot}$ for the
  batch are saved.
\item Batches are processed until we have generated the desired 
  number $N_t$ of target galaxies.  Note that each batch draws an
  independent set of templates.  The final shear estimate and its
  uncertainty are derived from the summed $P,\vecQ,\matR$ using
  Equations~(\ref{gbar}) and (\ref{cg}).
\end{enumerate}

\begin{deluxetable}{ccc}
\tablewidth{0pt}
\tablecolumns{3}
\tablecaption{Parameters and results of the baseline validation tests\label{sims}}
\tablehead{
\colhead{Characteristic} & 
\colhead{Gauss test} & 
\colhead{\galsim\ test}
}
\startdata
Galaxy profile & Gaussian & Decentered disk+bulge \\
PSF profile & $\delta$-function & Moffat, $\beta=3.5$ \\
PSF size (pixels) & \nodata & $r_{50}=1.5$ \\
PSF ellipticity & \nodata & $(0.00, 0.05)$ \\
Weight function & Gaussian & $k\sigma,$ eqn. (\ref{ksigma}) \\
Weight size & $\sigma=1$ & $\sigma=3.5$~pix \\[5pt]
Galaxy radius\tablenotemark{1} & 0.5--1.5 & 1.0--2.0 \\
Galaxy $S/N$ & 5--25 & 5--25 \\
$\sigma_e,$ galaxy shape noise & 0.2 & 0.2 \\
Selection cuts & none & $8<S/N<20$ \\[5pt]
$N_{\rm batch}$ / $N_{\rm template}$, target/templates per batch & 
$10^6$ / $3\times10^4$ & $5\times10^5$ / $2.5\times10^4$ \\
$\sigma_{\rm max}$ / $\sigma_{\rm step}$, template
truncation/replication &  5.5 / 1.0 & 6.0 / 1.1 \\
$N_{\rm sample},$ templates subsampled & $7\times10^4$& $5\times10^4$
\\[5pt]
$N_t$, total targets & $10^9$ & \edit{$10^9$} \\
Selection fraction & 1.0 & 0.69 \\
$\vecg_{\rm true},$ input shear & $(0.01, 0.00)$ & $(0.02, 0.00)$ \\
$(\vecg_{\rm meas}-\vecg_{\rm true})\times10^5$ & $(+0.1, +0.0)\pm(0.4,0.4)$ &
\edit{$(+4.3,-1.3)\pm(0.9,0.9)$} \\
\edit{Non-linearity $\alpha$} & \edit{2} & \edit{0.5} \\
\enddata
\tablenotetext{1}{Galaxy half-light radius is given relative to the
  weight scale for Gauss tests, or relative to the PSF half-light
  radius for \galsim\ tests.}

\end{deluxetable}

\subsection{Gauss tests}
We use the analytic moments of the Gauss tests to check the BFD
formulae and their implementation, and explore the sampling parameters
of the integration algorithm.  Table~\ref{sims} describes the baseline
simulation; in Section~\ref{approx} we investigate dependence of shear
bias on these parameters using the Gauss tests.
Although the moment calculations
are analytic, we use the full $k$-d tree implementation described in Section~\ref{kdtree} to 
evaluate the integrals.  We can quickly run a sufficient number of statistics to reach the
accuracy of $m\sim10^{-3}$ using these analytic simulations.  

Galaxy moments $\vecM^G$ (and their shear derivatives) are calculated
analytically, and the moment noise $\vecM^n$ is generated from the
multivariate Gaussian distribution with the known $\matC_M.$ A
complication is that the moment noise is held fixed as we shift the
target coordinate origin to null the \vecX\ moments.  This is contrary
to the behavior of normal images, and results in some changes to the
formulae for $P(\vecD | \vecg)$ which are described in
Appendix~\ref{translationnoise}. The baseline Gauss test with $10^9$ targets yields 
$m=(+0.1\pm0.4) \times10^{-3}.$

\subsection{\galsim\ tests}
\label{galsimsec}
\begin{figure}[ht]
\centering
\mbox{\subfigure{\includegraphics[width=3.4in]{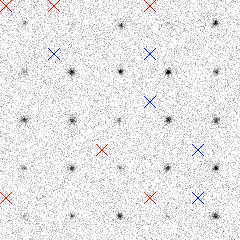}}\quad
\subfigure{\includegraphics[width=3.4in]{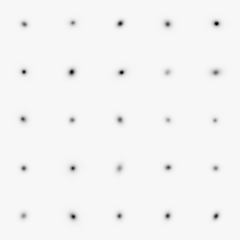} }}

\caption{ A sample of the target (left) and template (right) simulated
  galaxies used in the validation test. Targets are marked here with
  an X in the upper left of their stamp if they were cut for low (red)
  or high (blue) $S/N$.  } 
\label{galsim_images}
\end{figure}

The \galsim\ tests validate several aspects of the code that are not
exercised in the Gauss tests, primarily the measurement of moments and 
PSFs from pixelized images.  The \galsim\ code is used to produce
FITS images, each consisting of $100\times100$ postage stamps that are 
$48\times48$ pixels in size. Every stamp
contains one galaxy located near its center. Each galaxy is the sum of
an exponential disk and a deVaucouleurs bulge. Both components are
given the same ellipticity and half-light radius. The fraction of flux
in the bulge component is uniformly distributed between 0 and 1.
The center of
the bulge is randomly shifted with respect to the center of the disk
by a distance up to the half-light radius. 
For target galaxies, we apply a lensing shear \vecg.
We convolve the final galaxy with an elliptical Moffat PSF. 
If the galaxies are being used as targets,
Gaussian noise is applied to the final stamp image. 
A selection of targets and templates is shown in Figure~\ref{galsim_images}.
 
The range of flux assigned to galaxies is set such that it yields
$5<S/N<25$ for a circular galaxy of typical size under
matched-aperture detection.   In measuring shear, we
set selection bounds $f_{\rm min}=8\sigma_f, f_{\rm max}=20\sigma_f$.
Note that the selection uses a different definition of $S/N$ than the
generation.  At fixed flux, the selection favors more compact and more
circular galaxies.

The properties for these simulated galaxies were chosen 
to capture the non-idealities of real data which might affect the BFD
implementation:
\begin{itemize}
\item We give the PSF an ellipticity $e_{2}=0.05$, which will test our
  ability to reject PSF asymmetries. 
\item The Moffat PSF is not strictly band-limited so the data are
  slightly aliased.  The PSF half-light radius of 1.5 pixels yields a
  sampling equivalent to \emph{DES} imaging in seeing with FWHM of
  0\farcs8, which would be in the worst-sampled quartile of the data.
\item The decentering of the disk and bulge components breaks the
perfect elliptical symmetry of the galaxies, which might otherwise be
canceling some systematic error in the method.
\item Elliptical Gaussians are a six-parameter family, and hence a
  given point in the 6d $(\vecM^G,\vecX^G)$ space has only a single
  possible value for the shear derivatives.  
  The varying bulge fraction and bulge/disk misregistration in the
  \galsim\ simulations admit a 
  range of shear derivatives at each point in moment space.
\item These tests include a non-trivial selection function and hence
  test the validity of the BFD terms for non-selection.
\end{itemize}

We produce a total of \edit{$N_t=8.6\times10^8$ targets,} of which a fraction
0.69105 pass the flux selection test.  The calculated $P(s)$ from
\eqq{ps1} predicts this extremely well: $0.69111\pm0.00006$.  The
uncertainty on $P(s)$ arises from sampling noise in the template set.

Most importantly, the inferred values for $g_1$ and $g_2$ imply
\begin{align}
\label{results}
m & =\edit{(+2.1\pm0.4)\times10^{-3},} \\
c & =  \edit{(-1.3\pm0.9)\times10^{-5}.}
\end{align}

\edit{We detect a deviation from $m=0$ at $5\sigma$ significance: 
well below that demonstrated by any
  previous practical method.  The $c$ value is within $1.5\sigma$ of
  zero, and suppresses the input PSF ellipticity by a factor of $>3000.$}

If we omit the selection terms in
Equations~(\ref{qtot2}) and (\ref{rtot2}), we obtain \edit{$m=-0.0122\pm0.0004.$}
The selection term is clearly necessary for
part-per-thousand shear inference, and the BFD formalism appears to
calculate the correction to 20\% accuracy or better.

Lastly we can assess the accuracy of the code's internal estimates of
the uncertainty on the shear estimator.  The standard deviation of the
$g$ components derived from each batch of targets is
\edit{$(3.56\pm0.04)\times10^{-4},$  consistent with the internal error
estimate from \eqq{cg} of $3.58\times10^{-4}.$}

\section{Testing approximations}
\label{approx}
We collect here all the assumptions and approximations that have been made
in deriving the lensing inference formulae:
\begin{enumerate}
\item We have implicitly assumed that we know $\tilde I(\veck)$ at all
  values of $\veck$ with non-vanishing $\tilde T(\veck)$,  in other
  words that we have a Nyquist-sampled real-space image.
\label{nyquistitem}

\item The pixel noise $\vecD^n$ is stationary and independent of the
  underlying galaxy $G$, and the moment noise likelihood is a
  multivariate Gaussian.
\label{bgnoiseitem}

\item The Jacobian determinant $J = |d\vecX/d\vecx_0|$ 
  is positive at any location where there is
  non-negligible probability of selection
\label{convexitem}

\item Galaxies are uncrowded, in that
  no other galaxies contribute significantly to $\vecM$ or $\vecX$ at
  any location $\vecx$ where galaxy $G$ might be selected.
\label{uncrowdeditem}

\item The lensing is weak, so that
  a second-order Taylor expansion about $\vecg=0$ captures all
  information about $P(\vecM | \vecg).$
\label{weakitem}

\item Our template set $G$ is a complete sample of source galaxies.
\label{sampleitem}

\item We have a noiseless, unlensed image of each template.
\label{noiselessitem}

\end{enumerate}
In this Section we will describe our progress to date in verifying
that failures of these assumptions or approximations will not stand in
the way of achieving part-per-thousand inference of \vecg. 
\edit{Also: our \galsim\ simulation results, while very good, are still
imperfect, with $m$ measured $5\sigma$ deviant from zero.  So we are
interested in whether any of these approximations could be responsible
for this deviation.}

\subsection{Nyquist sampling}
\label{nyquistsec}
We have implicitly assumed Nyquist sampling of the data by defining
our moments as integrals over the regions of $\veck$ space with
non-vanishing $W(|k^2|).$  We will not in this paper examine the
consequences of aliasing in the data due to finite sampling.  We do
note, however, that the method does not require that the data be
available at all \veck\ or even that it be free of aliasing.  We can
\emph{define} our \vecM\ elements to be sums over a finite sampling of \veck\
space, and the formalism remains valid as long as we know what the
template galaxies' $\vecM^G$ would be under the same sampling, and
also know the first two derivatives of $\vecM^G$ with respect to
lensing distortion $\vecg$.  This is true even in the case of
aliasing, as long as the templates are aliased in the same way as the
targets.  This, however, is hard to arrange in practice, and it is
better to construct un-aliased data from dithered images if necessary,
as described in a simple case by \citet{Lauer} and in a more general
case by \citet{imcom}.  

Note also that the Moffat PSF used in the
validation tests of Section~\ref{validation} is not strictly
band-limited, so these tests incurred a level of aliasing that would
be typical for a well-designed ground-based survey.  \edit{In future tests
we will evaluate whether this aliasing, or some other approximation in
the \galsim\ rendering, is causing the non-zero $m$ value in the
\galsim\ tests.}

\subsection{Stationary noise}
The assumption of stationary, source-independent noise is valid for
background-limited (or read-noise-limited) imaging, which will
generally be the case for the galaxies dominating the lensing
information in ground-based weak-lensing surveys.  
We leave for future work the investigation of the impact of shot noise
from source photons, which may be relevant for low-background space-based surveys.
\edit{Our current simulations do not include source shot noise.}

\subsection{Convex galaxies}
\label{convexsec}
The assumption of positive Jacobian determinant $J$ for all
selectable regions was necessary to render as feasible the analytic
integration of selection probability, and also to avoid calculating
the joint probability of multiple detections of the same source.  
We consider two potential modes of failure of this approximation.

First, if the galaxy is sufficiently faint, $J^G$ is small enough even
near its peak that the noise in $\vecM$ can flip the sign and create
a fold in the $\vecx\rightarrow\vecX$ mapping. Clearly the defense against this is to
have the selection threshold $f_{\rm min}$ be large enough
($\gtrsim5\sigma_f$) that $J^G$ is also large enough to dominate the
noise fluctuations.  Further work is needed to determine if there is a
level of $f_{\rm min}$ which satisfactorily suppresses noise detections
without discarding sources that carry a significant fraction of the
lensing information.

Second, there will be galaxies which have high flux but have complex
structure such that $J^G$ crosses or approaches zero because of
multiple maxima or plateaus.   We should note that we only
care about the structure in the galaxy \emph{after} it has been
smoothed by the detection filter, which in real space is the Fourier
transform of $W(|k^2|)/\tilde T(\veck).$  We will usually aim to have
$W\approx \tilde T^2 \tilde I^g,$ where $\tilde I^g(\veck)$ is the transform
of the average observed galaxy.  Thus in practice, the observed image,
already convolved by the PSF, is convolved again with the PSF and the
typical galaxy profile before running the detection scheme.  This
means that any maxima or plateaus on scales of the PSF or smaller are
going to be erased.

We have not yet validated BFD on sources
with well-resolved structure that might lead to multiple selections,
but our implementation includes some ameliorative 
measures in anticipation of the issue.  First, in our postage-stamp
tests, we can discard all but the highest-flux detection in each
stamp.  Our calculation of $P(\vecM,s|G, \vecg)$ should then incorporate the
probability of a higher-flux selection existing.  Our crude version of
this is to include in our sum over $\vecu$ only those values of $\vecu$
that have $J^G\ge0$ and are contiguous with global maximum for $f$.
In other words we assume that the brightest detection will arise from
the convex region of the (filtered) galaxy that surrounds the global
maximum.  We have not yet quantified the efficacy of this approach on
realistic galaxies.

\subsection{Uncrowded galaxies}
Overlapping galaxies pose a considerable challenge for BFD (and indeed
for nearly all lensing-measurement methods).  We have strived for a
formalism that makes minimal assumptions about the morphology of the
galaxies.  But galaxy image deblending depends fundamentally on having
some prior expectations for galaxy morphology in order to partition
the flux in a single pixel among two (or more) sources.  We suspect
that for mild cases of blending, one could precede the BFD analysis
with joint model-fitting to multiple overlapping sources; and then,
subtract each source model in turn when measuring the Fourier moments
of the other.  This would likely be successful as long as the
subtracted flux has moments that are small compared to the remainder,
as our dependence on the correctness of the model will remain weak.
At present, we will simply ignore galaxies which overlap to an
extent that they grossly perturb each other's moments.  Crowding
remains as a critical issue for deep ground-based surveys, where the
product of typical observed galaxy size and desired target number
density is $\gtrsim 0.1$ \citep{chihway}.

\subsection{Weak lensing limit}
\label{weaklimit}
Any shear estimator that is analytic in the input shear and introduces no
preferred direction on the sky should have 
\begin{equation}
\label{nonlin}
\left\langle \vecg_{\rm meas} - \vecg_{\rm true} \right\rangle = \left[m +
\alpha g^2 + O(g^4)\right] \vecg_{\rm true}.
\end{equation}
\edit{The coefficient $\alpha$ is expected to be of order unity unless
$d\log P/d\vecg$ becomes large for some targets.  This will occur only
for galaxies whose moments are many $\sigma$ different from any of the
templates, a situation we avoid by adding noise to high-$S/N$
targets.  If $\alpha\sim 1$, then the desired accuracy of $<10^{-3}$
of the shear will be lost in
the second order Taylor expansion about $\vecg=0$ for
$g \gtrsim 0.03$.}
Expanding around $g \ne 0$
greatly complicates the calculation of the moments and their derivatives, making
it impractical.  BA14 derive third-order expressions, but these are
not included in our present implementation and would also slow the
method substantially.

\begin{figure}[ht]
\centering
\includegraphics[width=3.2in]{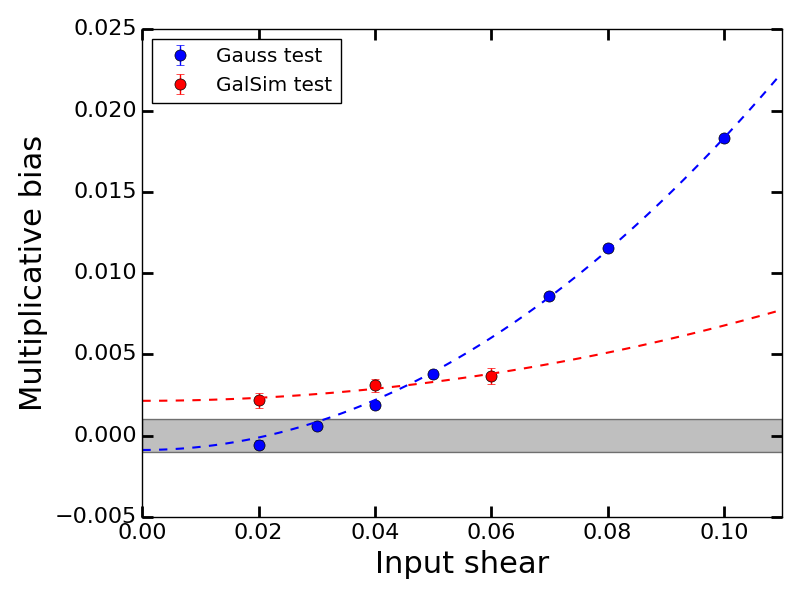}
\caption{ The recovered multiplicative bias using the second-order BFD
  formalism, as a function of
input shear.  The grey band shows the 
desired accuracy of $m<10^{-3}$.  The dashed line is a fit of the data to the
expected quadratic dependence on $g,$ with coefficient $\alpha\approx
2$ for the Gauss tests and $\alpha\approx0.5$ for the \galsim\ tests.}
\label{bias_shear}
\end{figure}

Figure~\ref{bias_shear} shows the recovered multiplicative bias as a
function of input shear in tests for non-linear behavior of both the
Gauss and \galsim\ tests.
The bias is consistent with the expected quadratic growth with $g$,  
with $\alpha\approx2$ and $\approx 0.5$ in
the two case.  A similar result is obtained for a model-fitting
implementation of the BA14 by \citet{ngmix}.
The value of $\alpha$ clearly can vary based on the nature 
of the galaxies and the noise levels.  
\edit{The $\alpha=0.5$ nonlinearity contributes
  an apparent multiplicative error of $\alpha g^2=0.2\times10^{-3}$ to
  our principal \galsim\ test results, smaller than the measurement
  error even with $10^9$ target galaxies.  But if real cosmic-shear
  measurements have a value closer to the $\alpha=2$ seen in the Gauss
  tests, the nonlinearity cannot be ignored:
The cosmic-shear test is, at its most basic, a measure of the
RMS dispersion $\sigma_g$ of the point distribution function (PDF) of shear to
$z\sim1$ sources.  Propagating (\ref{nonlin}) through a
nearly-Gaussian PDF suggests that we would mis-estimate $\sigma_g$ by
a factor $1+3\alpha\sigma_g^2,$ which for $\sigma_g\approx0.02$ and
$\alpha=2$ would be a fractional error of 0.0024 on $\sigma_g,$
larger than the expected statistical error 
for future surveys, and in need of a correction. Fortunately the value
of $\alpha$ is straightforwardly assessed to the $\sim20\%$ accuracy
that would be needed to render nonlinearity errors negligible.}

The nonlinearity poses a potentially larger problem for regions of high shear 
such as galaxy clusters. 
This problem can be overcome with an iterative procedure if we are
fitting a model to the shear. We first fit the model, ignoring
nonlinear shear response.  In regions of not-so-weak shear, \eg\ where
the shear is $\approx0.1,$ we can unshear the source galaxy by the predicted 0.1
shear before measuring its moments and calculating
$P_i, \vecQ_i,$ and $\matR_i$ with the nominal second order
procedure.  This will yield a Taylor expansion of $P(\vecD_i | \vecg)$
of deviations from the model, which can then be used to refine the model.
\edit{We speculate that}
this procedure would recover full unbiased accuracy around
well-measured individual clusters.

\subsection{Noiseless, unlensed templates}

\begin{figure}[ht]
\plottwo{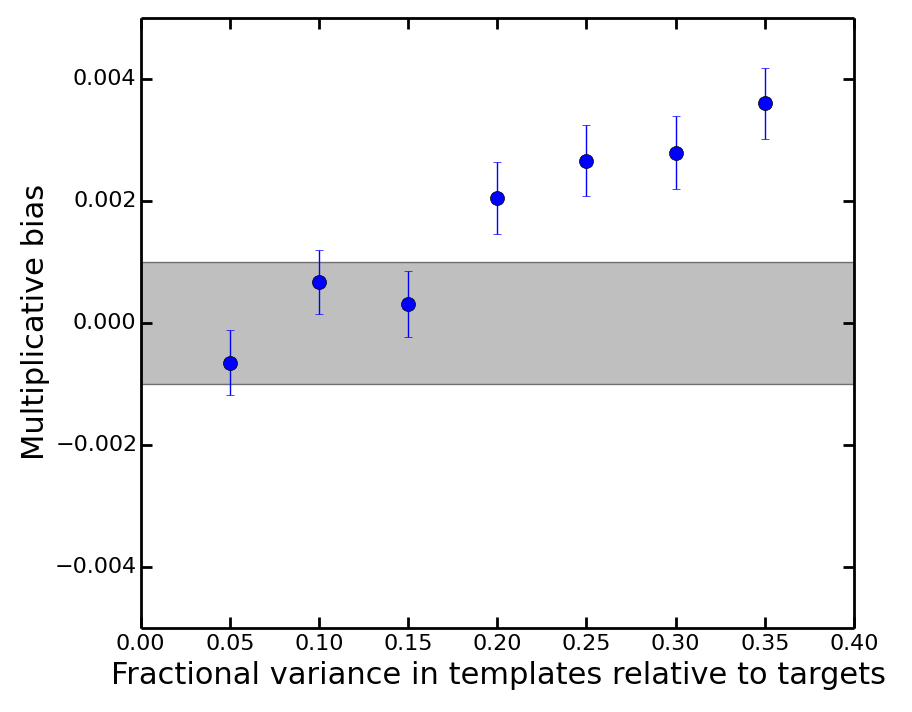}{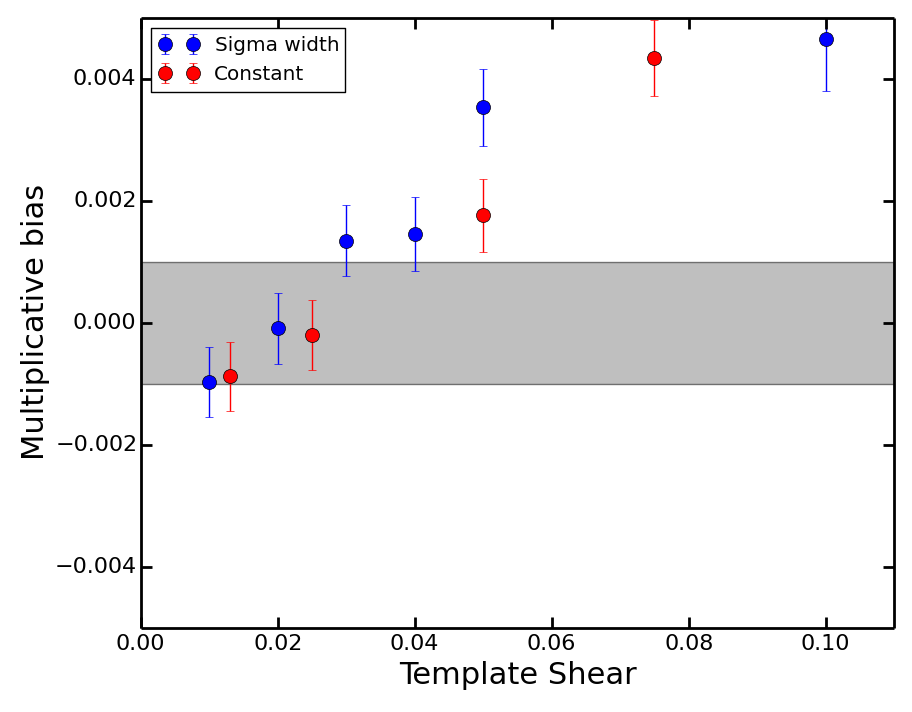}
\caption[]{At left: the multiplicative bias in shear inferred from
  simulations where  we add noise to the template
galaxies.  The $x$ axis gives the noise variance on the templates
relative to the noise variance on the target observations.  At right:
the multiplicative bias recovered when template galaxies
  have an applied shear that is either constant or drawn from a
  Gaussian of given RMS value.  The grey band shows the 
desired accuracy of $m<10^{-3},$ which we see is retained when the
templates have $\lesssim10\%$ of the noise power of the targets, and
the RMS shear on templates is $\lesssim0.04.$}
\label{template_noise}
\end{figure}

We have assumed that the galaxies used in constructing the prior are noiseless
and unlensed, whereas in real data they will be both noisy and lensed by large scale
structure.  We run a series of Gaussian simulations to evaluate
the impact of each of these on shear measurement.  In the first test, we add
noise to the moments and derivatives of the template galaxies.  Figure~\ref{template_noise} shows
the bias in recovered shear as a function of the ratio of template
noise variance to target noise variance.  When this ratio is
$\sim 10\%,$ the 
multiplicative bias remains $<10^{-3}.$  Therefore, observations with
$>10\times$ the integration time of target galaxies are sufficiently
high-$S/N$ to use as ``noiseless'' template galaxies.  Most
current and future lensing surveys already include such deep
observations to maximize their scientific value.

In the second test, we assess the impact of using template galaxies
that have already been sheared.  Recall that we use rotated copies of
all observed templates, which means that the \emph{mean} shear on our
templates is always zero, but we must ask whether non-vanishing \emph{variance}
of the shear on the templates produces biased shear inferences.
We applied shear to the templates in two different 
ways: a constant shear amplitude for every template; and a shear
randomly drawn from a zero mean Gaussian with dispersion $\sigma_g.$
Fig~\ref{template_noise}
shows the bias as a function of the applied template shear.  The
multiplicative bias satisfies $|m|<10^{-3}$ in both cases
if the RMS template shear is $<0.04$.  
\edit{The typical shear imparted by large scale structure is only half
  as large, thus it
appears feasible to use deep integrations of the real sky to produce
the template set.}

\subsection{Complete template set}
\label{samplesec}
\begin{figure}[ht]
\plottwo{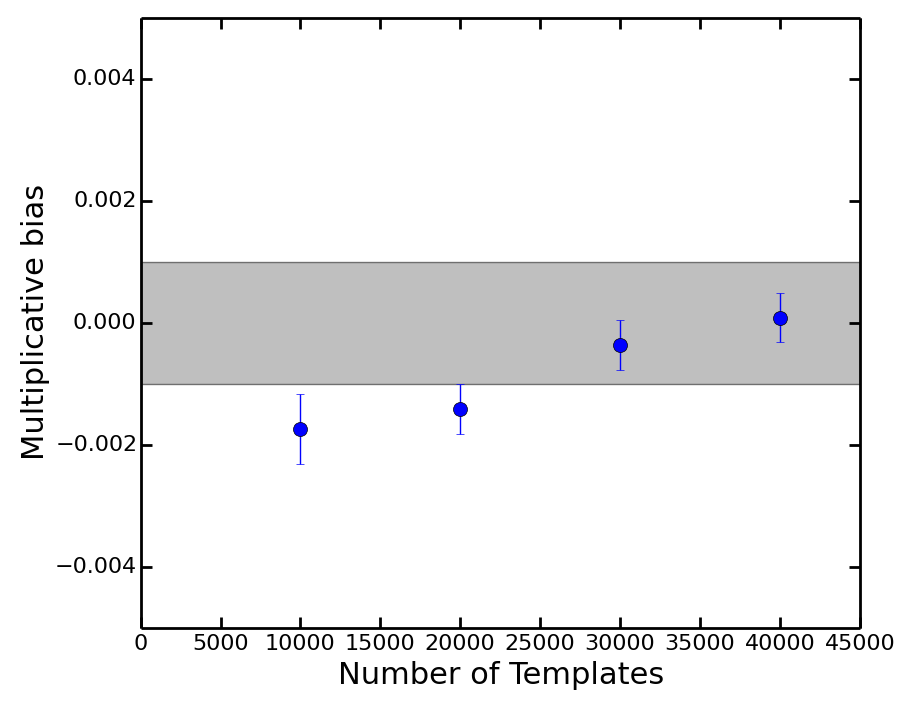}{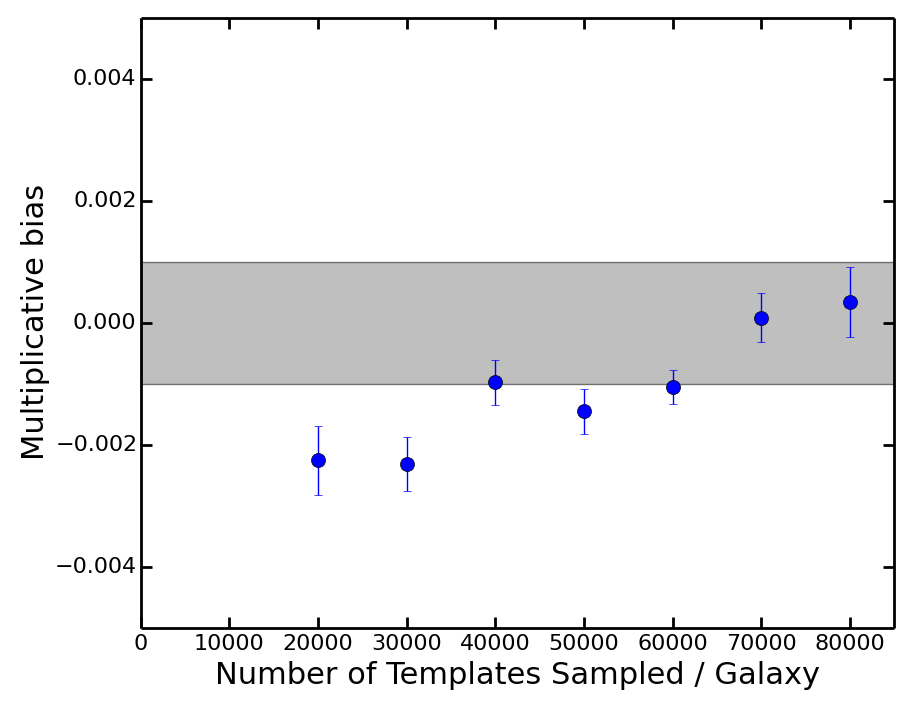}
\caption{ The multiplicative bias recovered as a function of the number of
  initial template galaxies before adding rotated/translated copies (left) and as a function
  of $N_{\rm sample}$ (right), the number of subsampled templates used to
  evaluate the integrals for each target galaxy.
  The grey band shows the desired accuracy of $|m|<10^{-3}$.
}
\label{template_n}
\end{figure}

We approximate the integral over all possible template galaxy types
and locations with a finite number $N_{\rm template}$ of high-$S/N$
galaxies, and by using a finite number of copies of each at 
intervals of $\sigma_{\rm step}$ in translation and rotation.  Further
we subsample a number $N_{\rm sample}$ of the resultant copies that
lie within $\sigma_{\rm max}$, of each target.
Ideally, the variance due to these approximations will be far below 
the expected noise of the targets. BA14 suggest that BFD estimates
will have a bias that scales inversely with the number of templates. 
We again use the Gauss tests to 
see how sensitive the recovered shear is to $N_{\rm template}$ and
$N_{\rm sample}.$ 
Figure~\ref{template_n} shows that, for the Gauss tests, the bias is within
$|m|<10^{-3}$ 
when we have at least 30,000 template galaxies and $N_{\rm sample}
\gtrsim 70,000$. The necessary values will in practice depend on how
the galaxies are distributed in moment space, how noisy they are, and
how we implement the added-noise strategy of Section~\ref{templates}.
These tests suggest, though, that a sample of $10^4$--$10^5$ deep sky
templates will suffice.  This is readily attainable in all planned surveys.

\begin{figure}[htb]
\centering
\includegraphics[width=0.5\linewidth]{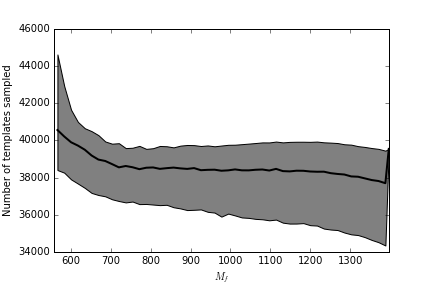}
\caption{ The number of templates used in the integration of
  $P(\vecD_i|\vecg)$ for each galaxy as a function of its flux moment
  $M_f$ in the \galsim\ tests.  The central line is the median,
  the shaded region bounds the 10--90 percentile range.}
\label{ntemp}
\end{figure}

\edit{The \galsim\ tests reported in Section~\ref{validation} used a value
of $N_{\rm template}=2.5\times10^4.$ Tests with a $2.5\times$ smaller value show no
significant change in $m$, arguing against the hypothesis that our
non-zero $m$ value is attributable to insufficient template
sampling. Figure~\ref{ntemp} shows that for the \galsim\ tests, we
integrate over a similar number of templates ($\approx 40,000$) for
each target.  Note that in this case $N_{\rm samp}$ was 50,000, so we
never invoked the subsampling of accessible templates.  An unrealistic
aspect of this test is that galaxies are uniformly distributed
in flux, leading to uniformity in the number of templates.  The real
sky has fewer bright galaxies and we would expect the template count
to increase for fainter sources.
}

\section{Future developments}
\label{extensions}
The BFD formalism can be straightforwardly extended beyond the basic
single-image, single-plane shear-estimation implementation that we
test in this paper.  In this section we sketch some of these
possibilities.

\subsection{Interferometric data}
\label{interferometry}
Interferometric data is collected in Fourier domain; data for a galaxy
will consist of estimates of $\tilde I(\veck)$ (the visibilities) at a
finite sampling of $\veck$ values determined by the interferometer
baselines.  As noted in Section~\ref{nyquistsec}, it is not required
that we measure galaxies at all \veck: we can
replace the integrals over $d^2k$ in \eqq{moments} with
weighted sums over the
visibilities.  The sole requirement is that we be able to calculate
the same sums for the template galaxies, as well as derivatives under
lensing distortion.   

\subsection{Multi-image analysis and multi-band data}
\eqq{moments} defines our compressed measurement vector \vecM\ as
derived from a single image $I(\vecx)$ observed with a single PSF
$T(\vecx)$.  In most surveys, observations of a given target will be
spread over multiple exposures $i\in\{1,2,\ldots,N\}.$ As long as each
individual exposure is unaliased, we can define the moments of the
target as a weighted sum over the moments $\vecM_i$ measured on each
exposure:
\begin{equation}
\label{weighted}
\vecM = \sum_i w_i \vecM_i.
\end{equation}
(and likewise for the detection moments \vecX).  It is important that
the $w_i$ be determined independent of the 
observed properties of the galaxy, so that the \edit{probability
$P(\vecM | G)$} remains calculable.  This linear combination of
the $\vecM_i$ yields a zero-mean normal distribution for the moment
noise, with covariance matrix $\matC_M$ that is the
sum of those for the individual exposures weighted by $w_i^2.$
Our implementation of this
extension selects the weights to minimize the variance of the
ellipticity moment $M_+$, a process which depends only on the noise
level and PSF of each exposure.

Formally, we can choose a different weighting function $W(|k^2|)$ for
each exposure, as long as we can calculate the $\vecM^G$ that would
result for each template (and its lensing derivatives).  
It is, however, convenient to use the same $W$ for all exposures, simplifying
construction of the template set.
If the seeing conditions of the exposures vary widely, then a single
$W$ may be far from optimum for some exposures; but since poor-seeing
exposures carry less lensing information to begin with, we lose little
by selecting $W$ to optimize the use of exposures at median or better seeing.

Because we make use of un-normalized moments, it is important that all
exposures (including templates) be placed on a common photometric
scale.

\subsubsection{Multiple observing bands}
\label{multiband}
There is also no requirement that all exposures be taken through the
same filter.  \eqq{weighted} can refer to exposures in multiple
filters.  We must once again select weights in
advance---iterative procedures such as weighting each galaxy according to 
its observed colors result in $P(\vecM | G)$ functions
that are analytically intractable.  Choosing fixed weights to apply to
each filter is akin to measuring moments in a bandpass that is
the weighted sum of all the filters' bandpasses.

Another alternative would be to define the moment vector \vecM\ to be
the concatenation of moment vectors from each filter.  This retains
more information, though at the cost of higher memory and 
computation demands due to the higher-dimensional moment space.  
We advocate a hybrid procedure, in which we retain distinct flux
moments $M_{f,i}$ for each filter $i$, but retain only a single
weighted combination of the other moments $M_r, M_+, M_\times, \vecX.$
This is because shape information is generally highly degenerate
between bands \citep{jarvisfilters}, but colors carry a lot of
information.  For example, red galaxies have a more compact shape
distribution than blue at low redshift \citep{bj02}, so retaining
color information when we compress the pixel data allows the BFD
formalism to exploit this distinction for more precise shear
inference.

Again the key requirement is that a low-noise measure of the template moments
$\vecM^G$ be available, as is the case if the templates are observed
in all of the same bands as the targets.  One can select distinct $W$
functions for each band, as long as the targets and templates are
treated consistently.  An advantage of a fixed $W$ across bands is
that the resultant flux moments then have the same pre-seeing window
function on the galaxy in all bands.  This property, also attainable with
PSF-matching codes, or the {\sc GaaP} algorithm of \citet{gaap}, is desirable
for use with photometric redshifts, since it insures that the measured
colors correspond to a fixed weighting of the stellar populations in
the galaxy, \ie\ we are not mixing aperture effects with stellar
evolution.

\subsection{Star-galaxy discrimination}
Stars are Dirac $\delta$ functions in real space, so their moments
$\vecM^G$ are known functions of flux and position $\vecx_G$.  Furthermore they are
unaffected by cosmological-scale lensing so we set the lensing derivatives of
$\vecM^G$ to zero.  If we add stellar sources to our template set,
assigning them $p_G$ values according to a prior expected sky 
density vs flux, then we automatically correct the shear estimate for
dilution by stellar sources.  We obtain as a by-product an excellent
posterior estimate of the probability that each source is stellar, and
we can sum these to obtain a posterior
stellar density estimate which may help to refine the stellar model that led to
the prior.  

It should also be noted that faint galaxy targets which might be confused with
stars are by definition weakly resolved, and contribute very little to
$\vecQ_{\rm tot}$ and $\matR_{\rm tot}$ for shear estimation.  Hence
the shear estimation will have low sensitivity to mis-estimation of
the stellar density in the prior.

If we have observed in multiple filters and retained flux moments
$M_{f,i}$ in each band as described in Section~\ref{multiband}, we can
(and must) produce stellar templates across the color-magnitude
diagram, with $p_G$ values expressing the expected density vs color
and magnitude.

\subsection{Magnification}
The BFD formalism makes no assumptions about the nature of the lensing
distortion vector \vecg, except that we can simulate its action on
each template, and that $P(\vecD|\vecg)$ is well
approximated by a quadratic Taylor expansion.  This means that we can
include magnification $\mu$ along with the shear components $g_1$ and
$g_2$ with essentially no change except to increment the dimension of
the \vecQ\ and \matR\ derivatives.

\citet{HuffGraves} note that early-type galaxies define a narrow plane
in the space of flux, size, and concentration, which then enables
much-enhanced determination of magnification.  Our current BFD
implementation would not exploit this gain since our compressed data
vector lacks information on concentration.  This could be remedied by
adding a $|k^4|$ moment to \vecM.  Furthermore we would need color
information, \ie\ the series of flux moments proposed in Section~\ref{multiband},
to distinguish red galaxies in a desired redshift range.  There is no
need to ``teach'' BFD about the 
existence of the Huff-Graves relation.  Any such relation that exists
will be automatically exploited in the lensing constraints, as long as
the action of lensing produces a shift in the way the galaxies
populate the moment space.

Two minor technical points about the estimation of magnification:
first, \eqq{pqrpoisson}  has
assumed that galaxies are placed by a Poisson process.  Clustering of
sources will mean that the resultant posterior is invalid,
underestimating the uncertainty on $\mu.$  
We can, if desired, treat the unlensed density $n$ as a spatial
variable when inferring magnification statistics, to distinguish
clustering from magnification.
Second, recall that magnification will dilute the source population on
the sky, changing the apparent $n.$  This effect is already included
in our implementation because we calculate the derivative
$\partial\vecM^G/\partial\mu$ by magnifying about the coordinate
origin, not the center of the galaxy.  This means that our grid
$\vecu$ of template copies is dilated by magnification, but we do
not alter $\Delta^2u$ in \eqq{pMs1} or \eqq{ps1}.  Thus source
dilation is included in the $P$ terms, and the $n$ term should retain
the unlensed density.

 Higher-order lensing distortions, \ie\ flexions, can similarly be
constrained by the BFD method, again as long as we augment the
compressed data vector to include quantities that are altered at first
order by the distortion.  Since flexion is not an affine
transformation, its action on the Fourier domain $\tilde I(\veck)$ is
less easily expressed than shear and magnification.  Nonetheless, it is
possible to derive flexion derivatives of template moments for
simultaneous constraint of all these lensing distortions using BFD.

\subsection{Lensing tomography and photometric redshifts}
One important caveat to BFD is that one cannot select subsets of
the targets and then combine their $P_i,\vecQ_i,\matR_i$ values to
estimate the shear on this subset.  This would
invalidate the $P(\vecD | \vecg)$ formulae we have derived, unless
one can guarantee that the {\it post hoc} selection criteria do not at
all alter the distribution of underlying moments $\vecM^G$ of the
selected galaxies.

Many useful scientific inferences and diagnostic tests for weak
lensing measurements rely upon comparing \vecg\ on
subpopulations of the sources.  Most critically, the bulk of the
lensing information, plus constraints on contamination by
intrinsic galaxy alignments, require splitting the source population
into redshift bins, a.k.a. lensing tomography.

If precise redshift estimates are available for all targets and
templates, then the application of BFD is straightforward, as we
compare each target only to templates that reside in the same redshift
bin.  More commonly we have \emph{probabilistic} redshift estimates
for targets derived from photometric redshift (\photoz) estimation.
Partitioning target or template galaxies by their maximum-likelihood redshifts will
not, in general, yield valid BFD inferences on the shear in each redshift
bin.\footnote{The same is true
  for most other lensing-inference methods: the responsivities or
  empirical calibrations they employ will depend upon the galaxy
  selection in subtle ways that may thwart part-per-thousand calibration.}

This apparent stumbling block turns out to be an opportunity: the BFD
formalism contains within it an ideal Bayesian \photoz\ estimation
mechanism, particularly for the sources with modest
$S/N\lesssim30$ photometry that dominate the weak lensing information
in most surveys. \citet{BPZ} presents the formalism for Bayesian
inference of redshift from broad-band fluxes; like BFD, it relies upon
having noiseless data vectors for a sample of ``truth'' objects of
known prevalence on the night sky.

We generalize the BFD method as follows: the lensing vector is
extended to the tomographic information
$\vect=\{\vecg_1,\ldots,\vecg_Z\},$ where $\vecg_\nu$ is the 
lensing distortion applied to sources in redshift bin $\nu$ out of $Z$
total bins.  We want the posterior $P(\vect | \vecD),$ and as before
we compress the image data $\vecD$ into the moments $\vecM_i$ of
objects detected and selected at positions $\vecx_i$.  The posterior
on \vect\ is calculable once we have expressions for $P(\vecM_i,s |
\vect)$ and the total selection probability $P(s | \vect).$
If we know the probability $p_{G\nu}$ that
template galaxy $G$ is in redshift bin $\nu,$ then we have the clear
generalization of \eqq{pMs1} to the tomographic case:
\begin{align}
\label{tomo1}
P(\vecM, s | \vect) & = \sum_{G,\vecx_G} p_G \, \Delta^2u\sum_\nu 
p_{G\nu} P(\vecM, s | G, \vecu,\vecg_\nu)\\
P(\vecM, s | G, \vecu, \vecg_\nu) & = \left| J(\vecM) \right|
\likeli\left[\vecX^G(\vecu,\vecg_\nu) \right]
\likeli\left[\vecM- \vecM^G(\vecu,\vecg_\nu) \right]
\end{align}
In the second line we make explicit the dependence of the template
moments on its true position $\vecu$ relative to the detection
location and upon the shear $\vecg_\nu$ to the source. The
total probability of detection vs \vect\ is similarly obtained
by introducing $p_{G\nu}$ into \eqq{ps1}.

We can also easily calculate the posterior redshift distribution $P(\nu |
\vecM_i)$ for each source, which would be found equivalent to the
treatment of \citet{BPZ}.  Of course our redshift discrimination will
be weak unless we have measured flux moments $M_{f,j}$ in multiple
bands $j$ as described in Section~\ref{multiband}, where we noted that
our pre-seeing aperture-matched fluxes are ideal for \photoz\
purposes.  We also note that we are working with fluxes, not colors,
and therefore we automatically include the ``luminosity prior'' that
is often added by hand into \photoz\ estimators. Indeed our inclusion
of $M_r,M_+,$ and $M_\times$  means that we automatically exploit any
size, surface brightness, or ellipticity information that helps with
redshift discrimination. 

Extending BFD to return the full tomographic lensing
likelihood $P(\vect | \vecD)$ would have many advantages for precision
lensing cosmology.  It would allow us to extract all the available
lensing information from 
galaxies with low-resolution \photoz\ information due either to color
ambiguities or low $S/N.$  It eliminates the need for {\it post hoc} estimation of
selection biases induced on the lensing estimators by \photoz\ cuts.
\edit{An important issue will be whether it is feasible to use
  sufficiently many, narrow bins that we do not need to worry about
  the variation of lensing signal across the redshift range of a bin.}

The implementation of BFD tomography requires that we have a template
galaxy set with known redshift probabilities assigned to each. 
Clearly one issue is how to obtain this information---especially since
the number of templates required to sample the moment space with
desired density will increase substantially with additional flux
dimensions in \vecM.  
It is likely infeasible to obtain spectroscopic redshifts for a
sufficiently large and complete set of templates.
A survey such as \emph{DES} which observes galaxies in the {\it grizY}
bands requires higher-$S/N$ observations in these bands to create the
template moments set; these observations in the survey bands could be
supplemented with deep data in other bands and with other instruments to tighten
the $p_{G\nu}$ estimates for the template set---for example, the
COSMOS field has data in many bands across the EM spectrum, producing
much higher-reliability \photoz's beyond the spectroscopic limit
\citep{ilbert}.  It will likely be necessary to use spectral-synthesis
methods to create artificially redshifted copies of the observed
template galaxies, just as we synthesize $\vecM^G$ for
rotated copies, 
in order to more densely sample the template space and damp the
line-of-sight structures
(sample variance)
present in the template fields. We envision BFD primarily as a means to
rigorously bootstrap the \photoz\ calibration from a well-observed subset of
galaxies to the full survey population.

Incompleteness in the spectroscopic surveys defining the redshift
priors is a difficult problem \citep[see \eg][]{BH08}.  The BFD tomography
formalism allows us to propagate the uncertainties due to missed
redshifts into the final cosmological results: we can reassign the
probabilities $p_{G\nu}$ using different assumptions about the missing
redshifts, and propagate these cases through
$P(\vect, \vecD)$ into cosmological inferences to determine their
impact.  One could also add a ``mystery bin'' to \vect\ to which we
assign all template galaxies with poorly known redshift.  The
tomographic BFD formalism will calculate the likelihood that any given
target has unknown $z,$ and cosmological inferences could marginalize
over the redshift distribution of the mystery bin.

Joint BFD tomography and \photoz\ is clearly an intriguing and
critical extension of the method, with quite a few details to work
out.  We will examine these in future publications.

\section{Conclusion}

The BFD method is now a practical, validated means to estimate WL shear at
parts-per-thousand accuracy.  \edit{In our initial large-scale tests, deviations from
perfection were measurable only with trials of nearly $10^9$ simulated
galaxies.  Work remains to determine if and why BFD has inaccuracies
at the level of $m=0.002.$ Future work will investigate the possible
impact of aliasing, and approximations in the rendering of images onto
finite postage stamps, since we find $m$ consistent with zero for our
Gauss tests that do no rendering.}
PSF asymmetries are perfectly removed from the shear estimator, to
present accuracy.
We have implemented a flux selection in such a way that we can
correct the $>1\%$ selection bias induced on
the shear.
The BFD formalism needs no parameter tuning or calibration to
eliminate biases---there is just a free weighting function one chooses to
minimize noise.  A real implementation does have some parameters for
sampling the infinite distribution of galaxies on the sky, which imply
a tradeoff of bias vs observational and computational resources.
We have used simulations to
validate the performance of the method and demonstrate that the
desired accuracy is attainable with readily available resources to
sample the underlying galaxy population.

While BFD assumes that noiseless images of unlensed galaxy populations
are available, our tests indicate that it retains desired accuracy
when the templates are taken from images with the same instrument, but
$\approx10\times$ longer exposure time than the target survey.  This
is eminently practical, and indeed most planned surveys already have
such ``deep fields'' for other reasons.  

The BFD method also predicts the uncertainty on the shear
estimate, and the detection efficiency, correctly to within the shot
noise of our tests.  The algorithms should scale to the needs of
even the largest proposed surveys, and the computational steps are
simple, highly parallel and amenable to execution on GPU's if greater
speed is needed.

\citet{ngmix} reports $|m|<2\times10^{-3}$ when applying the BA14
formalism to likelihoods derived from MCMC model-fitting to galaxy
images with $S/N$ as low as 10.  The galaxy images were both drawn and
fitted with simple 
S\'{e}rsic models, so this work notes that the method may be susceptible
to ``model bias'' in more realistic cases.  \cite{ngmix} also does not
yet include a prescription for galaxy selection and resultant biases.
 
The only other demonstration of part-per-thousand WL inference at
$S/N\lesssim10$ from a
realistic algorithm of which we are aware is \citet{zlf}, also
implemented as the {\sc FourierQuad} method in the \great\ challenge
\citep{great3}. This method shares several characteristics with BFD:
galaxies are reduced to weighted moments in Fourier space, where PSF
correction is straightforward.  Neither method assigns shapes to
individual galaxies; {\sc FourierQuad} works by stacking un-normalized
moments of the power spectrum; the shear estimator is a quotient of
stacks.  Using the power spectrum has the advantage of making the 
estimator insensitive to choice of galaxy origin, but amplifies
measurement noise by $\sqrt{2}$ relative to our phase-sensitive moments.  More
problematic is that a stacking method weights galaxies by flux,
which is far from optimal.  {\sc FourierQuad} does not yet have an
approach to selection and weighting of sources without biasing shear
inferences.  BFD is at this time closer to applicability on real data.

\citet{Schneider} propose an ambitious effort to simultaneously model
the shear field, the pixel-level appearance of galaxies within it, and
the underlying distribution of the source galaxies.  This approach
shares some formalism with BFD, but does ultimately rely on parametric
models for the galaxies.  \edit{Our ``model,'' which is that galaxies'
  true moments are equal to those of galaxies found in a deep sub-survey,
 should be less subject to model bias than the \citet{Schneider} approach
while greatly reducing the computational complexity.}

There are issues to address before BFD can be applied to real survey
data.  Working in Fourier space means we cannot easily exclude
pixel data contaminated by cosmic rays or defects, and hence we need
some method for infill of pixels or rejection of exposures.
Overlapping or multi-peaked galaxy images are not handled by BFD, so
we will need some combination of model-based deblending with rejection
of hopeless overlaps that does not significantly bias WL
inferences.  This will be easier in low-density surveys such
as \emph{DES} and \emph{KiDS} than deep ground-based surveys such as
\emph{LSST} and \emph{HSC}.  For space-based surveys, we need to
investigate the behavior of BFD in the presence of source shot noise
that violates our background-limited (stationary) noise assumption.
We also may need to develop a nonlinearity correction for some applications.

Our validation tests assume constant shear across all galaxies, but as
BA14 point out, it is straightforward to calculate a posterior
likelihood on the parameters of any model of shear vs position, for
example for tangential shear vs radius around a selected lens
population.
Cosmological models, however, predict a power spectrum or
other statistical property of the WL field rather than predicting the
shear pattern itself.  Current 2-point (and 3-point) estimators for
shear assume that each source galaxy provides a point estimate of the
shear, but BFD returns a different kind of information, namely some
weak probability distribution for shear along each line of sight in the
form of $\{P_i, \vecQ_i, \matR_i\}.$ 
Exploitation of the BFD outputs for lensing statistics will require
development of new estimation frameworks.  \citet{MMSS} discuss means
to treat such outputs as point estimators, and quadratic estimators
for 2-point functions that use BFD-style information.

We have also treated the lensing distortion as a single screen,
whereas the sources are distributed in $z$ and hence we measure a
weighted mean of shear on the line of sight.  A real experiment will
need to estimate the $z$ distribution of sources---or more precisely,
the distribution of contribution to the BFD shear estimate.  Better
yet, we have outlined an extension of BFD to joint Bayesian redshift
and shear estimation, which directly generates a tomographic lensing
likelihood $P(\vecD_i | \vect)$ for each source where \vect\ contains
the shear (and potentially magnification and source density) at a
series of $z$ bins.  This could open the door to full exploitation of
the low-to-modest $S/N$ regime---where both \photoz\ and WL estimators
have proven difficult to produce without bias---that potentially carries more
information than high-$S/N$ galaxies with well-constrained \photoz's.  Work is 
needed to develop statistics to constrain cosmological models with this
$P(\vecD_i|\vect)$ information, as opposed to the binned point estimates used
now.  It is likely that there are extensions of the \citet{MMSS}
techniques to this tomographic case.

A critical question will be
how many template galaxies must be observed, particularly in the
tomographic case where we will need to increase the dimensionality of
the moment space that the templates sample.  This is related to the
question of how large and complete a spectroscopic sample is needed to
calibrate \photoz's to the accuracy needed for WL cosmology.

The BFD method also naturally extends to multi-filter or
interferometric observations, and deals gracefully with the blurring
of the stellar and galactic loci in faint surveys.  Compared to
currently dominant model-fitting methods for WL inference, BFD has some
disadvantages, such as not-quite-optimal use of the pixel information,
annoyances with defective pixels, and a less-clear route to using
crowded sources.  BFD's advantages are, however, substantial, primarily in the superior accuracy
that comes from having a first-principles treatment of noise and
selection, and no need to assume a functional form for the sources.

\acknowledgments
We thank Michael Jarvis, Rachel Mandelbaum, and Barney Rowe for
leading the development of the excellent \galsim\ package, and thank
them plus Erin Sheldon for their advice on the BFD work.  \edit{Our
  referee Joe Zuntz improved this paper as well.} This work
was supported by grants AST-1311924 from the National Science Foundation,
DE-SC007901 from the Department of Energy, NNX11AI25G from NASA, and
cooperative agreement with JPL under NASA grant ROSES-12-EUCLID12-0004.
RA acknowledges financial support from Princeton University.

\newpage
\appendix
\section{Probabilities for augmented moment noise}
\label{addnoise}
Section \ref{templates} describes a strategy of adding noise $\vecM^A$
with covariance matrix $\matC_A$ to the moments $\vecM$ of a selected
target galaxy.  We need to know the probability of selecting the
galaxy and obtaining the total moments ${\mathcal M}=\vecM +
\vecM_A$.  Recall that the originally measured moments can also be
expressed as $\vecM = \vecM^G + \vecM^n,$ where the noise moments have
known covariance matrix $\matC_M$.  We assume that both $\vecM^n$ and
$\vecM^A$ are drawn from zero-mean multivariate Gaussians.

The galaxy is detected by the criterion $\vecX = \vecX^G + \vecX^n=0$
and selected according to $f_1 < M_f < f_2,$ and we will use the
notation $\vecM \in S$ to denote when this condition is satisfied.
We want the quantity
\begin{align}
P({\mathcal M},s | G) & = \likeli(\vecX^G) \int_{\vecM \in S} d\vecM\, \left|
                        J(\vecM) \right| P({\mathcal M},\vecM |
                        G) \\
 & = \likeli(\vecX^G) \int_{(\vecM^n+\vecM^G) \in S} d\vecM^n\, \left(
   J^G + 2\vecM^G\cdot \matB \cdot \vecM^n + \vecM^n \cdot \matB \cdot
   \vecM^n\right)
   \likeli({\mathcal M}-\vecM^G,\vecM^n).
\label{addnoise1}
\end{align}
Recall that we are approximating that the Jacobian derivative
$J=|d\vecX/d\vecx_0|=\vecM\cdot \matB \cdot \vecM$ is positive wherever
the likelihood is non-negligible.  \edit{Also note that in this
  Appendix we will suppress the dependence of moments and other
  quantities on the lensing \vecg\ and the displacement \vecu.}

The joint distribution of the final and initial noise
$\likeli({\mathcal M}-\vecM^G,\vecM^n)$ is a zero-mean normal
distribution.  The covariance matrix of the concatenated noise vectors
is known and fully specifies the distribution:
\begin{align}
{\bf Cov}({\mathcal M}) & = \matC_A + \matC_M \equiv \matC \\
{\bf Cov}(\vecM) & = \matC_M \\
{\bf Cov}({\mathcal M}, \vecM) & = \matC_M.
\end{align}
\eqq{addnoise1} can be integrated over the Gaussian distribution; the
result is
\begin{align}
\label{addpMs}
P({\mathcal M},s | G) & = \likeli(\vecX^G) \Delta^2x\, \left|2\pi
  \matC\right|^{-1/2}
\exp\left[-\frac{1}{2}\left({\mathcal M}-\vecM^G\right)^T \matC^{-1} 
\left({\mathcal M}-\vecM^G\right)\right] \\
\nonumber & \phantom{=} \times \left\{
Y \left[  J(\tilde\vecM) + {\rm
            Tr}\left(\matB\matC_A\matC^{-1}\matC_M\right)\right]
- 2 Y^\prime \vecZ^T \matB \tilde\vecM
+ Y^{\prime\prime} \vecZ^T \matB \vecZ\right\}, \\
\tilde\vecM & \equiv \matC_M\matC^{-1}{\mathcal M} + \matC_A\matC^{-1}\vecM^G, \\
\vecZ & \equiv \frac{1}{\sigma_f} \matC_A \matC^{-1} \matC_{Mf}, \\
\sigma^2_f & \equiv \left(\matC_A\matC^{-1}\matC_M\right)_{ff}.
\end{align}
Here $\matC_{Mf}$ is row $f$ of the original moment covariance
matrix.  The bounded Gaussian integral over $f$ results in the terms
\begin{align}
Y = Y(\nu_{\rm min}, \nu_{\rm max})
 & \equiv \int_{\nu_{\rm min}}^{\nu_{\rm
                                 max}} d\nu\, e^{-\nu^2/2}, \\
Y^\prime & \equiv \frac{\partial Y}{\partial u_{\rm max}} - \frac{\partial Y}{\partial u_{\rm min}}, \\
Y^{\prime\prime} & \equiv \frac{\partial^2 Y}{\partial u_{\rm max}^2} 
- \frac{\partial^2 Y}{\partial u_{\rm min}^2}, \\
\nu_{\rm min} & = \frac{1}{\sigma_f} \left( f_{\rm min} - \tilde
              M_f\right) ,\\
\nu_{\rm max} & = \frac{1}{\sigma_f} \left( f_{\rm max} - \tilde
              M_f\right).
\end{align}
We replace \eqq{pMs1} with a weighted sum of (\ref{addpMs}) over
template galaxies $G$ and their potential displacements $\vecx_G.$  As
before, derivatives with respect to \vecg\ propagate through the
expressions into derivatives of the template moments.  The total
selection probability from \eqq{ps1} is unaltered, since $\vecM^A$ is
not added to the moments until the selection process is complete.

\section{Probabilities with translation-invariant noise vector}
\label{translationnoise}
In some of our validation tests, we create moment vectors $\vecM$ for
targets by calculating $\vecM^G$
directly from analytic formulae rather than pixelated images, and adding
moment noise $\vecM^n$ drawn from its known distribution.  These simulated
targets differ from image-based simulations in that the moment noise
realization is invariant under shift of the coordinate origin $\vecx_0.$ In this
case the Jacobian $J=d\vecX/d\vecx_0$ has contributions only from the
underlying galaxy, not from the noise.  We therefore must alter our
formulae in this case of translation-invariant moment noise
realizations.  Equations~(\ref{pMs1}) and (\ref{ps1}) are altered by
substituting these equations for the moment probability and the total
selection probability of each galaxy:
\begin{align}
P(\vecM, s | G) & = \likeli\left(\vecX^G\right) \Delta^2x\,
                  J\left(\vecM^G\right)\,\likeli\left(\vecM-\vecM^G\right) \\
P(s | G) & = \likeli\left(\vecX^G\right)  \Delta^2x \, J\left(\vecM^G\right)\,  Y, 
\end{align}
where $Y$ is defined in \eqq{gaussY}, \edit{and we suppress dependence on
\vecg\ and \vecu.}

In the case where we have added noise to the moments after selection
in order to better smooth the template samples, we replace the formula
(\ref{addpMs}) with the simpler
\begin{equation}
P({\mathcal M},s | G) = \likeli(\vecX^G) \Delta^2x\, \left|2\pi
  \matC\right|^{-1/2}
\exp\left[-\frac{1}{2}\left({\mathcal M}-\vecM^G\right)^T \matC^{-1} 
\left({\mathcal M}-\vecM^G\right)\right] \, J(\vecM^G).
\end{equation}

\section{Derivatives and transformations of the Fourier-domain moments}
\label{momentcalcs}
Implementation of the BFD method requires that we calculate the
derivatives of the Fourier-domain moments of our template galaxies
under lensing distortions.  We summarize here the formulae for these
derivatives in the case of shear.  We give the formulae in the case
where the observed template surface brightness $I(\vecx)$ is a
continuous function.  The transition to finite sampled data is
straightforward. 

Our convention for the Fourier transform of the image is
\begin{equation}
\tilde I(\veck) = \int d^2x\, I(\vecx) \exp(-i\veck\cdot\vecx).
\end{equation}
We are interested in the change in moments after the image undergoes
an affine transformation
\begin{equation}
I^\prime(\vecx) = I\left(\matA^{-1}\vecx-\vecx_0\right)
\end{equation}
Standard Fourier manipulations give
\begin{align}
\tilde I^\prime(\veck) & = |\matA| e^{-i\veck^\prime\cdot\vecx_0} \tilde
I\left(\veck^\prime\right) \\
\veck^\prime &\equiv \matA^T \veck.
\end{align}
We define a two-component shear $\vecg=(g_1,g_2)$ of a galaxy image
with the flux-conserving transformation 
\begin{equation}
\matA^{-1} = 
\frac{1}{\sqrt{1-g^2}} \left( \begin{array}{cc}
1 - g_1 & -g_2 \\
-g_2 & 1 + g_1
\end{array}
\right).
\end{equation}
Note the BFD method is agnostic about the definition of shear; this is
simply the choice for our implementation.

It is convenient to adopt a complex notation at this point:
\begin{align}
k & \equiv k_x + ik_y  &\partial & \equiv \frac{1}{2}\left(\frac{\partial}{\partial g_1} - i \frac{\partial}{\partial
  g_2}\right)  \notag \\
g & \equiv g_1 + i g_2 &
\bar \partial &\equiv \frac{1}{2}\left(\frac{\partial}{\partial g_1} + i \frac{\partial}{\partial
  g_2} \right)
\end{align}
With this notation the action of shear $\veck\rightarrow
\left(\matA^T\right)^{-1}\veck$ becomes
\begin{equation}
k \quad \rightarrow \quad k^\prime = \left( 1- g\bar g\right)^{-1/2} \left( k - g \bar
  k\right).
\label{shearcomplex}
\end{equation}

The moments we are interested in can also be compactly expressed in 
complex notation as well:
\begin{equation}
\label{momentcomplex}
M^\prime_\alpha  = \int d^2k\, \tilde I(k) W(k^\prime \bar k^\prime)
                  F_\alpha(k^\prime),
                \end{equation}
with
\begin{align}
  M_0 & = M_f & F_0 & =1 \notag \\
M_1 & = X_x + i X_y & F_1 & = ik \notag \\
M_2 & = M_+ + i M_\times & F_2 & = k^2 \notag \\
 & M_r & F_r & = k\bar k.
\end{align}
The shear derivative operators can be rewritten as
\begin{align}
\bnab_g & = \vecv \partial + \bar\vecv \bar \partial
& \vecv & \equiv \left(\begin{array}{c}
1 \\ i
\end{array}\right) \\
\bnab_g\bnab_g & = \vecv\vecv^T \partial^2 + \bar\vecv
\bar\vecv^T\bar \partial^2
+ 2 \matI_2 \partial \bar\partial
& \matI_2 & \equiv \left(\begin{array}{cc}
1 & 0 \\ 0 & 1
\end{array}\right)
\end{align}
Now the derivatives of the moments with respect to
shear are obtained by applying these operators to the moment
definition (\ref{momentcomplex}) after substituting in the shear wavevector
transformation (\ref{shearcomplex}).  For each moment, the derivatives
can be expressed as
\begin{align}
\bnab_g M_\alpha & = \int d^2k\, \tilde I(k) \left[ W(k\bar k)
  A_\alpha(k)  + W^\prime(k\bar k) B_\alpha(k) \right] \\
\bnab_g\bnab_g M_\alpha & = \int d^2k\, \tilde I(k) \left[ W(k\bar k)
  C_\alpha(k)  + W^\prime(k\bar k) D_\alpha(k) +
W^{\prime\prime}(k\bar k) E_\alpha(k) \right].
\label{mDerivs}
\end{align}
Table~\ref{mTable} lists the functions $A,B,C,D,E,$ and $F$ that
yield the moments and their derivatives under shear.
All of the moments and their derivatives are simple weighted
polynomial moments of the galaxy Fourier transform.

A translation of the galaxy by $\vecx_0$ adds a factor
$e^{-i\veck\cdot\vecx_0}$ to $\tilde I(\veck)$ in the integrand of all
the moments (and their derivatives).

\begin{deluxetable}{ccccc}
\tablewidth{0pt}
\tablecaption{
Functional forms of the integrands for moments and their derivatives,
as defined by Equations~(\ref{mDerivs}).  The derivatives of the
moments under translation in $x$ and $y$ directions are found by adding
factors of $i(k+\bar k)/2$ and $(k-\bar k)/2$ to the entries, respectively.
\label{mTable}
}
\tablehead{
\colhead{Moment} & \colhead{$M_0$} 
& \colhead{$M_1$} 
& \colhead{$M_2$} 
& \colhead{$M_r$} }
\startdata
$F_\alpha=$ & 
$1$ & $ik$ & $k^2$ & $k\bar k$ \\ \tableline

$A_\alpha = \vecv \times $ 
& 0 & $-i \bar k$ & $-2k\bar k$ & $-\bar k^2$ \\
\phantom{$A_\alpha$}$+\bar\vecv \times$ 
& 0 & 0 & 0  & $-k^2$ \\ \tableline

$B_\alpha = \vecv \times $
& $-\bar k^2$ & $-ik\bar k^2$ & $-k^2\bar k^2$ & -$k \bar k^3$ \\
\phantom{$B_\alpha$}$+\bar\vecv \times$ 
& $-k^2$ & $-ik^3$ & $-k^4$ & $-k^3 \bar k$ \\ \tableline

$C_\alpha = \matI_2 \times $ 
& 0 & $ik$ & $2k^2$ & $4k\bar k$ \\
\phantom{$C_\alpha$} $+\vecv \vecv^T \times $ 
& 0 & 0 & $2\bar k^2$ & 0 \\ \tableline

$D_\alpha = \matI_2 \times $ 
& $4k\bar k$ & $6ik^2\bar k$ & $ 8k^3 \bar k$ & $8k^2\bar k^2$ \\
\phantom{$D_\alpha$}$+ \vecv\vecv^T \times $ 
& 0 & $2i\bar k^3$ & $ 4k \bar k^3$ & $2\bar k^4$ \\
\phantom{$D_\alpha$}$+ \bar\vecv \bar\vecv^T \times $ 
& 0 & 0 & 0 & $2k^4$ \\ \tableline

$E_\alpha = \matI_2 \times $ 
& $2k^2\bar k^2$ & $2ik^3 \bar k^2$ & $2k^4 \bar k^2$ & $2k^3\bar k^3$ \\
\phantom{$E_\alpha$}$+ \vecv\vecv^T \times $ 
& $\bar k^4$ & $ik\bar k^4$ & $k^2\bar k^4$ & $k \bar k^5$ \\
\phantom{$E_\alpha$}$+ \bar\vecv \bar\vecv^T \times $ 
& $k^4$ & $ik^5$ & $k^6$ & $k^5 \bar k$
\enddata
\end{deluxetable}

\end{document}